\documentclass[prl,twocolumn,showpacs,preprintnumbers,amsmath,amssymb]{revtex4-1}

\usepackage{graphicx}
\usepackage{ulem}
\usepackage{xcolor}
\DeclareGraphicsExtensions{.png,.eps,.pdf}

\begin{document}
\title{ Quantum spin liquids in a square lattice subject to an Abelian flux and its experimental observation in cold atoms or photonic systems }
\author{ Fadi Sun and Jinwu Ye}
\affiliation{
Department of Physics and Astronomy,
Mississippi State University, Mississippi State, Mississippi 39762, USA  \\
Department of Physics, Capital Normal University,
Beijing, 100048, China    }
\date{\today }

\begin{abstract}
  We report that a possible $ Z_2 $ quantum spin liquid (QSL) can be observed in
  a new class of frustrated system: spinor bosons subject to a $ \pi $ flux in a square lattice.
  We construct a new class of Ginsburg-Landau (GL) type of effective action to classify possible quantum or topological phases
  at any coupling strengths.
  It can be used to reproduce the frustrated  SF with the 4 sublattice $ 90^{\circ} $ coplanar spin structure
  plus its excitations in the weak coupling limit and the FM Mott plus its excitations
  in the strong coupling limit achieved in our previous work. It also establishes deep and intrinsic connections between
  the GL effective action and the order from quantum disorder  (OFQD) phenomena in the weak coupling limit.
  Most importantly, it predicts two possible new phases at intermediate couplings: a FM SF phase or a frustrated magnetic Mott phase.
  We argue that the latter one is more likely and melts into a $ Z_2 $ quantum spin liquid (QSL) phase.
  If the heating issue can be under a reasonable control at intermediate couplings $ U/t \sim 1 $,
  the topological order of the $ Z_2 $ QSL  maybe uniquely probed by the current  cold atom or photonic experimental techniques.
\end{abstract}

\maketitle


{\bf 1. Introduction }
  It was well known that for a quantum anti-ferromagnetic Heisenberg (QAFH) model in a square lattice,
  the ground state is a quantum anti-ferromagnetic ( AFM) state \cite{aue,sachdev,frusrev}
  which breaks the spin $ SU(2) $ to $ U(1) $. However, due to the geometric frustration, the AFM state
  does not work for the QAFH in a triangular lattice.  In 1973,
  P. W. Anderson \cite{and1} suggested that the ground state could be a quantum spin liquid ( QSL) of
  the Valence bonds  which  does not develop a long-range magnetic order or any other orders even at zero temperature.
  In 1987, after the experimental discovery of the high temperature superconductors, he speculated that doping
  the QSL could lead to the high temperature superconductors \cite{and2}.
  Unfortunately, the ground state in a triangular lattice turns out to be magnetically  ordered with a 3 sub-lattice $ 120^{\circ} $
  co-planar state which completely breaks the spin $ SU(2) $ symmetry\cite{sachdev,frusrev,subir}.
  Even so, Anderson's idea sparked great interests to find the existence of QSL in other geometrically frustrated systems.
  For example, it does appear
  in the Rokhsha and Kivelson's quantum dimer model in a triangular lattice \cite{dimer1,dimer2,dimer3},
  may also appear in the QAFH model with $ J_1-J_2-J_3 $ interactions in a square lattice at least in the large N limit \cite{largeN,largeN2}.
  For QAFH in a Kagome lattice which provides much stronger quantum fluctuations than a triangular lattice,
  the ground state is likely to be a QSL whose nature remains controversial \cite{sachdev,frusrev,subir}.
  The 2d QSL maybe classified from the projective symmetry group (PSG) of its anyonic excitations \cite{wen1,wen2}.
  The topological features and long-range entanglement of the $ Z_2 $ QSL can be best seen in the exactly solvable
  model called "Toric code " \cite{Kit1} constructed by Kitaev in 2013:
  it has $ 2 \times 2 $ topological degeneracy in a torus and also a topological entanglement entropy (TEE) $ \log 2 $.
  The 2d Toric code can also be easily generalized to the 3d Toric code where the $ m $ particle is just a
  loop excitation with a tension which resembles a vortex line or loop excitation in a Type-II s-wave superconductors \cite{myown}.


  In 2006, Kitaev studied a quantum compass model in a honeycomb lattice which has spin-bond
correlated spin exchange interactions \cite{Kit2}, so it lacks a spin $ SU(2) $ symmtry.
It is exactly solvable and hosts the gapped $ Z_2 $ QSL phase in the Toric code, most importantly, also a gapless $ Z_2 $ QSL phase which,
in the presence of  a small Zeeman field, turns into a gapped QSL phase hosting non-Abelian excitations, chiral
Majorana fermion edge mode and quantized thermal Hall conductivity.
Then it was proposed \cite{HKmodel} that combining Heisenberg model with the Kitaev model
called Heisenberg-Kitaev (HK) model may describe
possible QSL phases in some 4d or 5d strongly correlated materials with strong spin-orbit couplings such as
Iridates or Osmates, namely, so called Kitaev materials. Unfortunately, so far,
only Zig-Zag commensurate phase or In-commensurate Skyrmion crystal (IC-SkX)  phases were observed experimentally,
no QSL phases have been found. For a complete picture on QSL, see recent reviews \cite{SLrev1,SLrev3,CSL}.
In a series of works \cite{rh,rhht,rafhm,rhh,devil}, the authors
studied a 2d Rotated Ferromagnetic/Anti-ferromagnetic Heisenberg model (RFHM/RAFHM) which can be written as
the Heisenberg-Kitaev-Dzyaloshinskii-Moriya (DM) form. In some SOC parameter regime, the RFHM
consists a dominant FM Kitaev term plus a small AFM Heisenberg term and a small DM term, so matches the
experimental parameters well in these so called Kitaev materials,  it indeed hosts the IC-SkX phase in the SOC parameter regime.
Because it is the DM term which breaks the parity, so
even a small DM term plays a crucial role in the formation of such IC-SkX phases.

Despite the sound establishment of the $ Z_2 $ QSL, its fractionalized excitations, topological orders and
the long-range entanglements in various concrete theoretical models, its possible existence in real materials remains
tantalizing and continues to be elusive. In this work, we demonstrate that it could appear in the new class of frustrated
system studied by the authors \cite{gold} which is in a partite lattice and spin $ SU(2) $ invariant:
pseudo-spin 1/2 spinor bosons or photons hopping in a square lattice subject to an Abelian flux
in the intermediate coupling regime. It can be realized in
simple, clean and easily tunable bosonic cold atom or photonic systems.
  For simplicity and also practical relevance to the cold atom or photon experiments,
  we focus on the most frustrated case with $ \alpha=\pi $ flux.
  In the previous work \cite{gold}, we found a frustrated SF ground state with the 4 sublattice $ 90^{\circ} $ coplanar spin structure
  plus its excitations in the weak coupling limit and the FM Mott plus its excitations  in the strong coupling limit.
  Here we focus on the intermediate coupling regime.
  Unfortunately, there is no controlled microscopic calculations at any intermediate couplings, one must take a different approach.

    Here we develop a symmetry-constrained phenomenological Ginsburg-Landau (GL) type of effective action to describe
    all the possible phases and phase transitions at any coupling strengths.
    Using the effective GL action, we are able to match all the results achieved both in the weak and strong couplings in \cite{gold},
    therefore establish intrinsic connections between the phenomenological parameters in the GL action
    and the bare parameters in the microscopic Hamiltonian Eq.\ref{piflux}.  Especially, for the very first time,
    we establish a deep connection between a symmetry based phenomenological
    theory in this work with the microscopic calculation on the effective potential generated by OFQD at weak coupling limit
    performed in \cite{gold}.
    Using the effective GL action, we also predict two possible new phases in the intermediate coupling regimes: a FM SF phase and a frustrated magnetic Mott phase with the 4 sublattice $ 90^{\circ} $ coplanar spin structure. We argue that the latter is the likely case.
    By contrasting with
    the  3 sublattice $ 120^{\circ} $ coplanar spin structure in a triangular lattice, we argue that
    the frustrated magnetic Mott phase is likely melt into a $ Z_2 $ QSL phase
    in the same class as that in a 2d toric code, so it supports fractionalized spinon excitations with the topological orders.
    Some possible analogy with solid $ ^{3} He $ films is also made.
    We also discuss some possible connections between our effective GL action with the $ 2+1 $ dimensional WZW model with
    and without a topological term.
    The combination of the  symmetry-based phenomenological approach in this work with
    the microscopic calculations on both weak coupling analysis and strong coupling expansion in \cite{gold}
    leads to a complete physical picture of the system in all coupling regimes which include both
    exotic symmetry broken states at weak/strong couplings and topological ordered states breaking no symmetries at intermediate couplings.
  Because the heating effects may still be under good control at intermediate couplings, armed with the ability to directly detect
  the topological orders of the putative QSL,
  the recent cold atom ( or  photonic ) experiments in an
  Abelian flux in an optical lattice ( or in a microwave cavity array ) could be
  a completely new class of system to search for still elusive QSL.
  As by-products, we also achieve some new results on two component spinor bosons with no flux and one
  component in the $ \pi $ flux and also stress their crucial differences than the present 2 component/$\pi $ flux problem.


   We study a pseudo-spin-$1/2$  Boson-Hubbard model in a $\pi$-flux on the square lattice described by:
\begin{align}
    \mathcal{H}=
	-t \sum_{\langle ij\rangle} e^{i A_{ij} }
	 b_{i\sigma}^\dagger b_{j\sigma}+h.c.
	+\frac{U}{2}\sum_{i} n^2_{i}
	-\mu\sum_{i} n_{i}
\label{piflux}
\end{align}
where $b_{i\sigma}$ are the boson annihilation operators on site-$i$, $ A_{ij} $ is the gauge fields putting on the links,
$n_i=n_{i\uparrow}+n_{i\downarrow}$ is the total number of bosons,
$\mu$ is the chemical potential.
In the following, we only focus on the spin $ SU(2) $ invariant interaction.
Here we study Eq.\ref{piflux} in all the coupling regimes
instead of just weak and strong couplings in \cite{gold}.


{\bf 2. The effective action to describe the transition from the weak coupling to the strong coupling  }

   In \cite{gold}, we did microscopic calculations based on the microscopic Hamiltonian Eq.\ref{piflux}
   in both weak coupling and strong coupling limit.
   In the weak coupling limit where the small parameter is $ U/t $,
   we found the frustrated SF ground state with the 4 sub-lattice $ 90^{\circ} $
   coplanar structure and the 4 linear gapless modes.
   In the strong coupling limit where the small parameter is $ t/U $, we found the FM Mott state with the ground state as the FM
   and 1 gapless quadratic FM mode.
   Here we take a completely different approach: construct symmetry based Ginsburg-Landau (GL) effective action
   to study the Hamiltonian Eq.\ref{piflux} at any couplings including the intermediate coupling regime $ U/t \sim 1 $.

   To do so, we  rewrite the spinor boson order parameter in the weak coupling analysis in \cite{gold} as:
\begin{align}
    \Psi_\mathbf{r}=(\eta_1\otimes \mathbf{z}_1)e^{-i\mathbf{K}\cdot\mathbf{r}}
	               +(\eta_2\otimes \mathbf{z}_2)e^{+i\mathbf{K}\cdot\mathbf{r}}
\label{GLpara}
\end{align}
    where the two newly defined spinors $ \mathbf{z}_a, a=1,2 $ contain both the charge and spin sectors.
    Note that we only expand the boson field operator in terms of the two minima of the kinetic energy
    without assuming any symmetry breaking \cite{etafluc}.

  The density and spin can be expressed in terms of the two spinors as:
\begin{align}
	\langle\Psi_0^A|\sigma|\Psi_0^A\rangle
	&= \vec{n}_1 + \vec{n}_2 \cos(2\mathbf{K}\cdot\mathbf{r})   \nonumber  \\
	\langle\Psi_0^B|\sigma|\Psi_0^B\rangle
	&= \vec{n}_1 + \vec{n}_3 \cos(2\mathbf{K}\cdot\mathbf{r})
\label{ABtwo}
\end{align}
 where the two sublattices $ A $ and $ B $  in Fig.1(a) are listed separately and $ \mathbf{r} $
 stands for the unit cell and the three 3-vectors are defined as
\begin{align}
	\vec{n}_1&=(\mathbf{z}_1^\dagger \sigma \mathbf{z}_1+\mathbf{z}_2^\dagger \sigma \mathbf{z}_2)  \nonumber   \\
	\vec{n}_2&=(\mathbf{z}_1^\dagger \sigma \mathbf{z}_2+\mathbf{z}_2^\dagger \sigma \mathbf{z}_1)   \nonumber  \\
	\vec{n}_3&=i(\mathbf{z}_1^\dagger \sigma \mathbf{z}_2-\mathbf{z}_2^\dagger \sigma \mathbf{z}_1)
\label{threevector}
\end{align}
  or equivalently and more intuitively  $  \vec{n}_1=(\mathbf{z}_1^\dagger \sigma \mathbf{z}_1+\mathbf{z}_2^\dagger \sigma \mathbf{z}_2),
  \mathbf{z}_1^\dagger \sigma \mathbf{z}_2=\vec{n}_2 +i \vec{n}_3,
  \mathbf{z}_2^\dagger \sigma \mathbf{z}_1=\vec{n}_2 -i \vec{n}_3 $.

 Eq.\ref{ABtwo} can be expressed in terms of a function of the lattice index-$i$ (in unit $\hbar/2$):
\begin{align}
	\vec{S}_i
	=\vec{n}_1+\frac{1}{2}(\vec{n}_2+\vec{n}_3)(-1)^{i_x}+\frac{1}{2}(\vec{n}_2-\vec{n}_3)(-1)^{i_y}
\label{threeordering}
\end{align}
The  mean-field ground-state  in \cite{gold} corresponds  to
$\vec{n}_1=n_0\cos2\phi(0,0,1), \vec{n}_2=n_0\sin2\phi(0,1,0), \vec{n}_3=n_0\sin2\phi(1,0,0)$.
Note that only $ \vec{n}_1 $ is a conserved quantity, while $  \vec{n}_2,  \vec{n}_3 $ are not.
If one applies a uniform, a staggered at $ (\pi,0) $ or   $ (0, \pi) $ Zeeman field, it will couple to
$ \vec{n}_1 $ and $  \vec{n}_2, \vec{n}_3 $ respectively. Note the absence of the ordering wavevector $ (\pi,\pi) $ in Eq.\ref{threeordering},
so if applying a Zeeman field along $ \hat{z} $ at $ (\pi,\pi) $, namely $ -h_z \sum_i (-1)^{(i_x+i_y)} S_{iz} $,
then it drops out of the continuum GL effective action Eq.\ref{general}.
Interestingly, this case corresponds to the right Abelian line in Rashba SOC and QAH case. It seems one can only use microscopic calculations
to discuss this case.

 One can check the total magnitude of the spin:
\begin{align}
	( \vec{S}_i )^2
	=\vec{n}_1^2+\frac{1}{2}(\vec{n}_2^2+\vec{n}_3^2)
	+\vec{n}_1\cdot(\vec{n}_2+\vec{n}_3)(-1)^{i_x}    \nonumber  \\
	+\vec{n}_1\cdot(\vec{n}_2-\vec{n}_3)(-1)^{i_y}
	+\frac{1}{2}(\vec{n}_2^2-\vec{n}_3^2)(-1)^{i_x+i_y}
\label{totals}
\end{align}
 which seems not uniform. This should not be worrisome.
 As shown in the appendix D, if setting the ground  state solution
 $  \mathbf{z}_1^\dagger \mathbf{z}_2=0 $, then  $ \vec{n}_1\cdot \vec{n}_2=0, \vec{n}_1 \cdot \vec{n}_3=0, \vec{n}_2^2=\vec{n}_3^2 $ and
 $ ( \vec{S}_i )^2= n $ becomes a constant. So the three vectors $ \vec{n}_1, \frac{1}{2}(\vec{n}_2+\vec{n}_3),
 \frac{1}{2}(\vec{n}_2-\vec{n}_3) $  at the three ordering vectors $ (0,0),(\pi,0), (0,\pi) $ respectively in Eq.\ref{threeordering}
 are orthogonal to each other at the ground  states.
 If setting the equal magnitude condition $  |\mathbf{z}_1|^2=| \mathbf{z}_2|^2 $, the expressions can be simplified further
 ( appendix D ).

 For the density distribution, one can set $\sigma= 1 $ in Eq.\ref{threevector} and
 obtain a uniform density at the ground  states.

    Note that the decomposition Eq.\ref{GLpara} only writes the boson field in terms of the two low energy degree of freedoms
    near the two minima of the Kinetic term in Eq.\ref{piflux}. It does not invole any symmetry breaking yet.
    So the effective action should keep the $ SU(2)_s \times U(1)_c $ symmetry, the Time-reversal symmetry $ {\cal T} $,
    the translational $ T_x, T_y $ satisfying the magnetic space group
     $ T_xT_y=T_y T_x \omega $ with $ \omega=-1 $ and square lattice point group symmetry such as
    the $ C_4 $ rotation $ R_{\pi/2} $ and the reflection with respect to $ x $ or $ y $ axis $ I_x $ and $ I_y $ of the microscopic Hamiltonian  Eq.\ref{piflux}.  In terms of the two spinors,  $ \mathbf{z}_a $, the most general form which respects all these symmetries is:
\begin{align}
    \mathcal{L}[\mathbf{z}_a   ]
	& =\mathbf{z}_a^\dagger\partial_\tau \mathbf{z}_a
	+v^2(\nabla\mathbf{z}_a^\dagger)\cdot(\nabla\mathbf{z}_a)
	+r(\mathbf{z}_a^\dagger\mathbf{z}_a)  \nonumber   \\
	& +u_1 (\mathbf{z}_a^\dagger\mathbf{z}_a)^2
	  +u_2 [(\mathbf{z}_a^\dagger\tau_{ab}^x\mathbf{z}_b)^2
	  +(\mathbf{z}_a^\dagger\tau_{ab}^y\mathbf{z}_b)^2]    \nonumber   \\
	& +w(\mathbf{z}_a^\dagger\tau_{ab}^z\mathbf{z}_b)^2 + \cdots
\label{general}
\end{align}
where $ \cdots $ means higher order derivative terms which may break the rotational space symmetries to
the square lattice point group symmetries and also higher orders in the order parameters, $ \tau^n, n=1,2,3 $
are 3 pseudo-spin matrices in the $ a=1,2 $ space,
the repeated indices are implicitly summed over,
i.e. $\mathbf{z}_a^\dagger\mathbf{z}_a
=\mathbf{z}_1^\dagger\mathbf{z}_1+\mathbf{z}_2^\dagger\mathbf{z}_2$
and $\mathbf{z}_a^\dagger\tau_{ab}^z\mathbf{z}_b
=\mathbf{z}_1^\dagger\mathbf{z}_1-\mathbf{z}_2^\dagger\mathbf{z}_2$.
In the following, for simplicity, we consider $ u_1=2u_2=u $ case which holds in the weak couplings.

In the effective action, all the possible $ SU(2)_s \times U(1)_c $ invariants can be written
in terms of the elements $ (\mathbf{z}_a^\dagger\tau_{ab}^n\mathbf{z}_b)^2 , n=0,1,2,3 $.
The two $ SU(2)_s $ spinors  $ \mathbf{z}_a, a=1,2 $ may imply $ SU(2)_{s1} \times  SU(2)_{s2} $ symmetry, but
the crossing term $ [(\mathbf{z}_a^\dagger\tau_{ab}^x\mathbf{z}_b)^2
	  +(\mathbf{z}_a^\dagger\tau_{ab}^y\mathbf{z}_b)^2] $ break it into just one $ SU(2)_s $.
The parameters $v^2$,  $u$ are always positive.

\begin{figure}[!htb]
\centering
    \includegraphics[width=0.8\linewidth]{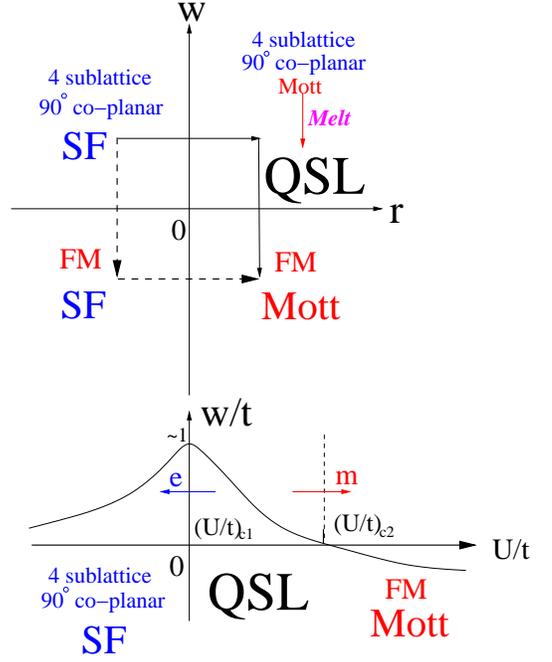}
	\caption{ Top: The 4 possible quantum or topological phases of the effective action Eq.\ref{general}
    in terms of its two phenomenological parameters $ r $ and $ w $. There are two possible scenarios
    (a) The frustrated SF to FM-SF to FM-Mott transition along the dashed line. The FM-SF was drawn in Fig.2c in \cite{gold}.
    (b)  The frustrated SF to a QSL to FM-Mott transition along the solid line.
    The frustrated SF and FM-Mott phases have been firmly established in the weak and strong coupling respectively. While the FM-SF and
    the $ Z_2 $ QSL are predicted by the GL effective action.
    Bottom: The expected  $ w $ dependence  on $ U/t $ is sketched at the bottom. $ r $ changes its sign at $ (U/t)_{c1} $.
    It is a monotonically increasing function of $ U/t $ in the SF phase, then starts to  decrease when entering the Mott side, then changes
    sign in the strong coupling limit. If so, then the solid line in (a) on the top is the likely case.
    The condensation of $ e $ ( or $ m $ ) particle leads to the frustrated SF ( FM Mott ) on the left (right ). }
\label{QSL}
\end{figure}

From the GL effective action, one can classify the 4 possible states as follows:
$ r < 0 $ is SF state,  $ r > 0 $ is Mott state.
$ w >0 $ Ising limit, $ w< 0 $ Ising limit.
The frustrated SF state at $ r < 0, w >0 $ matches that achieved in the weak coupling limit $ U/t \ll 1 $.
The FM Mott state at $ r  > 0, w  < 0 $ matches that achieved in the strong coupling limit  $ t/U \ll 1 $.
The two possible scenarios are sketched in Fig.\ref{QSL}.
The symmetry based effective action approach can not determine the sign of the $ w $ term
which can be  either positive or negative. $ w >0 $ means a easy-plane limit, while
$ w <0 $ means an Ising limit. In the following, we discuss the two cases separately.

{\bf 3. The easy plane limit  $ w >0 $: the frustrated SF and a possible $ Z_2 $ QSL
along the solid line in Fig.\ref{QSL}. }

{\sl (a) The frustrated SF at at $r<0$. }

 At $r<0$, the symmetry breaking pattern is $ SU(2)_s \times U(1)_c \rightarrow 1 $.
 In the easy-plane limit $w>0$,
 it is naturally to introduce a parametrization of $\mathbf{z}_1$ and $\mathbf{z}_2$ as
\begin{align}
	\mathbf{z}_\alpha=\sqrt{\rho_\alpha}e^{i\chi_\alpha}
	\begin{pmatrix}
		e^{-i\phi_\alpha/2}\cos(\theta_\alpha/2)\\
		e^{+i\phi_\alpha/2}\sin(\theta_\alpha/2)\\
	\end{pmatrix},
\label{zz}
\end{align}
then the mean-field grand-canonical potential density is
\begin{align}
	\Omega=& r(\rho_1+\rho_2)
	+u(\rho_1+\rho_2)^2
	+w(\rho_1-\rho_2)^2\nonumber\\
	&+2u\rho_1\rho_2\Big[\cos^2\Big(\frac{\phi_1-\phi_2}{2}\Big)
			    \cos^2\Big(\frac{\theta_1-\theta_2}{2}\Big)    \nonumber   \\
			   & +\sin^2\Big(\frac{\phi_1-\phi_2}{2}\Big)
			    \cos^2\Big(\frac{\theta_1+\theta_2}{2}\Big)\Big] + \cdots
\end{align}
which leads to a minimization condition
\begin{align}
	\rho_1=\rho_2=-\frac{r}{4u},\quad\phi_1-\phi_2=0,\quad\theta_1-\theta_2=\pi
\end{align}
Without loss of generality, we choose the saddle point solution  identical to that in \cite{gold} as
$\rho_1=\rho_2=\rho_0/2=-\frac{r}{4u}$,
$\chi_1=\chi_2=0$,
$\phi_1=\phi_2=0$,
$\theta_1=\theta_2+\pi=\pi/2$,
and expansion around the saddle point lead to
\begin{align}
    \mathcal{L}
	& =	i\delta\rho_\alpha\partial_\tau\delta\chi_\alpha +\frac{v^2\rho_0}{8}
	[\frac{1}{\rho^2_0}(\nabla\delta\rho_\alpha)^2 +4(\nabla\delta\chi_\alpha)^2 ]                     \nonumber   \\
    &+u(\delta\rho_1+\delta\rho_2)^2+w(\delta\rho_1-\delta\rho_2)^2   \nonumber   \\
	& +(-1)^\alpha\frac{i}{4}\rho_0\delta\phi_\alpha\partial_\tau\delta\theta_\alpha+
    \frac{v^2\rho^2_0}{8} [ (\nabla\delta\phi_\alpha)^2
	+(\nabla\delta\theta_\alpha)^2]
	\nonumber\\
	& +\frac{u\rho_0^2}{8}	[(\delta\theta_1-\delta\theta_2)^2 +(\delta\phi_1-\delta\phi_2)^2]
\label{eq:Hpi1}
\end{align}
    where the repeated indices is summed over $ \alpha=1,2 $ in Eq.\ref{zz}.

If one introduce
$\rho_\pm=\rho_{1}\pm\rho_{2}$,
$\chi_\pm=\chi_{1}\pm\chi_{2}$,
$\phi_\pm=\phi_{1}\pm\phi_{2}$,
and $\theta_\pm=\theta_{1}\pm\theta_{2}$,
then
\begin{align}
    \mathcal{L}_\text{GL}
	&=\frac{i}{2}\delta\rho_\alpha \partial_\tau\delta\chi_\alpha +
    \frac{v^2\rho_0}{4} [\frac{1}{\rho^2_0}(\nabla\delta\rho_\alpha)^2+(\nabla\delta\chi_\alpha)^2]
                             \nonumber   \\
    &+u(\delta\rho_+)^2+w(\delta\rho_-)^2    \nonumber   \\
	&+\frac{i}{8}\rho_0\delta\phi_\alpha \partial_\tau\delta\theta_{\bar{\alpha}}
	+\frac{v^2\rho_0}{16}
	[(\nabla\delta\phi_\alpha)^2
	+(\nabla\delta\theta_\alpha)^2]
	\nonumber\\
	& +\frac{u\rho_0^2}{8}[(\delta\theta_-)^2 +(\delta\phi_-)^2]
\label{eq:Hpi2}
\end{align}
   where the repeated indices is summed over $ \alpha=\pm $ and $ \bar{\alpha} $ means $ - \alpha $.

    From Eq.\ref{eq:Hpi2}, one can identify the four conjugate pairs $ ( \delta \rho_{\pm}, \delta\chi_{\pm} ) $
    and  $ ( \delta\phi_{\pm}, \delta\theta_{\mp} ) $ which lead to the two  eigen-modes $ \omega_{\pm} $ and
    $ \omega_2=\omega_3 $ respectively:
\begin{align}
	\omega_+=\sqrt{v^2k^2(v^2k^2+4\rho_0 u)}   \nonumber  \\
	\omega_2=\omega_3=\sqrt{v^2k^2(v^2k^2+2\rho_0 u)}   \nonumber \\
	\omega_-=\sqrt{v^2k^2(v^2k^2+4\rho_0 w)}
\label{fourleff}
\end{align}
  where $ \omega_{-} $ is dictated by $ w $, the other three are dictated by $ u $.

Obviously, all the 4 modes depend on $ \rho_0 $ explicitly and become unstable when $ r > 0 $.
$ \omega_+ $ is the SF mode, $ \omega_- $ is the spin mode generated by OFQD, $ \omega_{3,4} $ are the other two spin
modes.

{\sl (b)  A possible $ Z_2 $ quantum spin liquid at $r > 0$. }

When $ r >0 $, it becomes a Frustrated Mott state. It may be convenient to also separate the charge from
the spin sectors by writing  $ \mathbf{z}_1 \rightarrow \psi \mathbf{z}_1, \mathbf{z}_2 \rightarrow \psi \mathbf{z}_2 $
where $ \psi =\sqrt{\rho} e^{i \chi} $
is a complex field standing for the charge fluctuations and $ \mathbf{z}_1, \mathbf{z}_2 $ is two spinors satisfying
$ | \mathbf{z}_1 |^2 + | \mathbf{z}_2  |^2=1 $ standing for the spin fluctuations.
On paper, this could be a 4 sublattice $ 90^{\circ} $ coplanar spin structure Mott state in a square lattice.
In the appendix D, by making a detailed comparison with the 3-sublattice  $ 120^{\circ} $ coplanar
spin structure Mott state in a triangular lattice\cite{frusrev,sachdev}, we conclude this Mott state
is likey to melt into a $ Z_2 $ QSL phase with de-confined spinon $ z_{a} $ defined by:
\begin{equation}
 n_{2 \alpha} + i n_{3 \alpha} = \epsilon_{ac} z_c \sigma^{\alpha}_{ab} z_b
\label{spinon}
\end{equation}
 where $ |z_1|^2 + |z_2|^2=1 $. While the $ \vec{n}_1 $ quantum fluctuations in Eq.\ref{threevector} are so massive,
 that it can be dropped in the effective low energy $ Z_2 $ gauge theory.
 As to be argued below, the de-confined spinon $ z_{\alpha} $ are dramatically different than the
 original vectors $ \mathbf{z}_1, \mathbf{z}_2 $.

More intuitively, $ \mathbf{z}_1, \mathbf{z}_2 $ are spinors instead of just two complex numbers.
This may also due to the fact that the 4 sublattice $ 90^{\circ} $ coplanar spin structure Mott state in Eq.\ref{threeordering}
has three ordering wavevectors $ (\pi,0) $, $ (0, \pi ) $ and $ (0,0) $.
As demonstrated explicitly in the appendix D, there are two sources of quantum fluctuations
in the range $ (U/t)_{c1} < U/t < (U/t)_{c2} $ ( Fig.\ref{QSL} ), one out of the classically degenerate manifold  controlled by  $ U/t $,
another within the  classically degenerate manifold controlled by  $ w/t $,
which melt the putative frustrated Mott state into a $ Z_2 $ QSL
state which is sandwiched between the frustrated SF  at $ U/t < (U/t)_{c1} $ and the FM Mott state at  $ U/t > (U/t)_{c2} $.
Both of which can be considered as
the two different parent states of the $ Z_2 $ QSL and provide an ideal environment for the formation of the QSL.
Obviously, the deconfined spinons $ z_{\alpha} $ defined in Eq.\ref{spinon}  would be dramatically
different from the original  $ \mathbf{z}_1, \mathbf{z}_2 $
which are confined into $ \vec{n}_1, \vec{n}_2, \vec{n}_3 $ in Eq.\ref{threevector}.
So the  deconfined spinons ( also called $e $ particle carrying spin $ s=1/2 $ ) are not related to
$ \mathbf{z}_1, \mathbf{z}_2 $ in a direct way.
This phenomenon is similar to the unitarity puzzle \cite{stevekondo,Malkondo,kondoye12345,kondoye123452}
in multi-channel Kondo model where the
scattering Majorana  fermions are "orthogonal" to the original incoming Majorana fermions, so the scattering matrix between
the in-coming and out-going states becomes zero.
In the $ Z_2 $ QSL, there are also gapped $ m $ particle carrying $ s=0 $  standing for a vortex in the $ Z_2 $ gauge field.
Both $ e $ and $ m $ are bosons, but  there is a mutual semionic statistics \cite{myown} between the $ e $ and $ m $ particles, so their composite
$ \epsilon = em $ is a fermionic  particle with $ s=1/2 $. It also leads to the long-range entanglement
of the QSL with the topological EE $ \gamma= \ln 2 $.
All the possible 2d $ Z_2 $ QSL maybe classified by the projective symmetry group satisfied by its anyonic excitations \cite{wen2}.
For most recent review, see Ref.\cite{SLrev3}.

{\bf 4. The Ising limit $ w<0 $: FM SF and FM Mott state
along the dashed line in Fig.\ref{QSL}. }

 In the Ising limit $w<0$, 
  one can simply set $ \mathbf{z}_2=0 $ and
 $ \mathbf{z}_1=\mathbf{z} $. Then Eq.\ref{general} reduces to:
\begin{align}
    \mathcal{L}[z]
	&=\mathbf{z}^\dagger\partial_\tau \mathbf{z}
	+v^2(\nabla\mathbf{z}^\dagger)\cdot(\nabla\mathbf{z})
	+r(\mathbf{z}^\dagger\mathbf{z})   \nonumber  \\
	&+u(\mathbf{z}^\dagger\mathbf{z})^2+ \cdots
\label{singlecomponent}
\end{align}
which shows a quantum phase transition \cite{alsonoflux} from
a FM SF at $ r<0 $ to a FM Mott $ r>0 $.
Obviously, it is still has the same symmetry as Eq.\ref{general}.

 Let us call the $\rho= |\mathbf{z}|^2 $ which minimizes Eq.\ref{singlecomponent} as $\rho_0$:
\begin{align}
	\rho_0
	=\begin{cases}
	    0,&\text{ if }r>0,\\
	    -\frac{r}{2u},&\text{ if }r<0.\\
	\end{cases}
\end{align}

{\sl (a)  FM Superfluid at  $r<0$. }

When $r<0$, the bosonic system is in a FM superfluid phase with the symmetry breaking pattern
$ SU(2)_s \times U(1)_c \rightarrow U(1)_s $.
Without of loss generality, we choose the spin symmetry breaking along the X direction, so
choose the condensate as $(\rho,\chi,\theta,\phi)=(\rho_0,0,\pi/2,0)$.
The Langriagian Eq.\ref{singlecomponent} can be expanded as:
\begin{align}
    \mathcal{L}_\text{FM-SF}
	& =i\delta\rho\partial_\tau\delta\chi
	+\frac{i}{2}\rho_0\delta\phi\partial_\tau\delta\theta +u(\delta\rho)^2   \nonumber  \\
	& +\frac{v^2 \rho_0}{4}
	[\frac{1}{\rho_0^2}(\nabla\delta\rho)^2
	+4(\nabla\delta\chi)^2
	+(\nabla\delta\phi)^2
	+(\nabla\delta\theta)^2]  \nonumber  \\
	& +r\rho_0+u\rho_0^2
	+(r+2u\rho_0)\delta\rho + \cdots
\end{align}
where the linear term in the last line vanishes for $\rho_0=-r/(2u)$. Obviously, at the quadratic level,
the action can be decomposed into a charge part and a spin part  $ \mathcal{L}_\text{FM-SF}
	=\mathcal{L}_\text{FM-SF,c}+\mathcal{L}_\text{FM-SF,s}+ \mathcal{L}_{m} $ where
\begin{align}
    \mathcal{L}_\text{c}
	&=i\delta\rho\partial_\tau\delta\chi
	    +\frac{v^2}{4\rho_0}(\nabla\delta\rho)^2+\rho_0(\nabla\delta\chi)^2+u(\delta\rho)^2   \nonumber \\
    \mathcal{L}_\text{s}
	&=\frac{i}{2}\rho_0\delta\phi\partial_\tau\delta\theta
	    +\frac{v^2}{4}\rho_0[(\nabla\delta\phi)^2+(\nabla\delta\theta)^2]     \nonumber \\
    \mathcal{L}_m
	&=\frac{i}{2}\delta\rho\delta\phi\partial_\tau\delta\theta
	+\frac{v^2}{4}\delta\rho[(\nabla\delta\phi)^2+(\nabla\delta\theta)^2]   \nonumber \\
	&-v^2\rho_0\delta\theta(\nabla\delta\phi)(\nabla\delta\chi)
	+\cdots
\end{align}
   where $  \mathcal{L}_m $ stands for the cubic coupling between the charge and spin part.
   So in the presence of a $ \pi $ flux, the charge and spin are decoupled at the quadratic level.
   Just from the scaling analysis, one can see the cubic coupling term is irrelevant.

   From the above equation, one can easily extract one charge and one spin mode:
\begin{align}
    \omega_\text{c}
	&=\sqrt{v^2k^2(v^2k^2-r/2)}   \nonumber \\
    \omega_\text{s} &=v^2 k^2
\end{align}
    One can see that the charge mode becomes unstable when $ r > 0 $.
    However, the spin mode is independent of $ r $ and may remain un-critical through the QCP and
    reach the FM Mott phase.

{\sl (b)  FM Mott at  $r > 0$. }

When $ r >0 $, it becomes a FM Mott state with the symmetry breaking pattern
$ SU(2)_s \times U(1)_c \rightarrow U(1)_s \times U(1)_c $.
It is convenient to separate the charge and spin by writing $ \mathbf{z} \rightarrow \psi \mathbf{z}  $
where $ \psi =\sqrt{\rho} e^{i \chi} $
is a complex field standing for the charge fluctuations and $ \mathbf{z}=(z_1, z_2) $ is a spinor satisfying $ |\mathbf{z}|^2=1 $ standing for the spin fluctuations. Then the charge sector and spin sector are described by:
\begin{align}
    \mathcal{L}_\text{c}
	&= \psi^{\dagger} \partial_\tau \psi + t |\nabla \psi|^2 + r |\psi|^2 + u | \psi|^4 + \cdots  \nonumber   \\
    \mathcal{L}_\text{s}
	&=\mathbf{z}^{\dagger} \partial_{\tau} \mathbf{z} + v^2 ( \nabla \vec{n} )^2
\label{FMscaction}
\end{align}
 where $ \vec{n}= \mathbf{z}^{\dagger} \vec{\sigma} \mathbf{z}, \vec{n}^2=1 $ stands for a unit vector of the FM spin fluctuations.
 Then due to the particle-hole symmetry at the integer filling, the charge part can be written in the particle-hole
 excitations where it becomes a second derivative in $ \partial_{\tau} $.

{\bf 5. Compare GL action with the microscopic calculations at both weak and strong coupling:
        new physical interpretation of the OFQD at waek coupling }

 Comparing Eq.\ref{fourleff} achieved by the effective GL action with the four modes listed in \cite{gold} achieved by
 the microscopic calculation at weak coupling, we find they are the same  if identifying
 ( after unifying the notation $ \rho_0 = n_0 $ )
\begin{align}
 v^2 &= t/\sqrt{2},~~r=-(\mu+2 \sqrt{2} t)    \nonumber   \\
 u & = U/2,~~ w= B/(8n_0^2)
\label{mappingweak}
\end{align}
  where $ B \sim ( n_0 U)^2/t  $ was evaluated by the microscopic calculations in \cite{gold}.

 The second line in Eq.\ref{mappingweak} shows that the $ w $ term is nothing but the effective potential generated by OFQD.
 Indeed, when subsituting the parametrization of $\mathbf{z}_n$ used in \cite{gold} into
 the $ w $ term in Eq.\ref{general}, we obtain $w(\mathbf{z}_m^\dagger\tau_{mn}^z\mathbf{z}_n)^2
 =w n_0^2\cos^2(2\phi)=4w n_0^2(\delta\phi)^2$. This brings deep insights on the physical mechanism of the OFQD.
 Of course, the value of $ w $  deviates from this value when approaching the
 QCP at $ (U/t)_{c1} $ in Fig.\ref{QSL} from below. In fact, $ w $ changes its sign from the QSL to the FM Mott at  $ (U/t)_{c2} $:
 namely, the easy-plane limit changes to the Ising limit.

 When comparing Eq.\ref{FMscaction} achieved by the effective GL action with the results achieved
 in the strong coupling expansion in \cite{gold}, we conclude that in the strong coupling limit
\begin{equation}
 w=-4t^2/U<0,~~~  v^2=4t^2/U
\end{equation}
 which are very different from those at weak coupling in Eq.\ref{mappingweak}.

 However, because the charge sector has been projected out in the strong coupling expansion,
 so one will not be able to make connections between the phenomenological parameters in the first equation in
 Eq.\ref{FMscaction} with any microscopic calculation.
 Although there is no controlled microscopic calculations at the intermediate couplings $ t/U \sim 1 $,
 we can expect the behaviour of $ w $ in Fig.\ref{QSL}. If so, the solid line is the most likely the case.

 Unfortunately, any $ \pi $ flux effects do not show up in the leading order in the strong coupling in \cite{gold}.
 One need to get  to the next order of $ t^4/U^3 $ which includes ring exchange terms around
 a square to see the $ \pi $ flux.
 Let's call the resulting model as J-Q model.
 It may resemble the J-Q model describing the magnetism of solid $ He^3 $ mono-layer absorbed on graphite \cite{frusrev}
which also contains the FM $ J <0  $ term and a ring exchange $ Q > 0 $ term in a 4-sites plaquette.
The $ Q $ term differs in the sign between boson and fermion. However, here the $ \pi $ flux changes the sign back, so
the $ Q $ term  here has the same sign as that in solid $ He^3 $. It is this twist which may stabilize the QSL.
Of course, here, it is a square lattice, while the latter
is a triangular lattice, but the FM $ J $ term is in-sensitive to the underlying lattice,
while the $ Q $ term is on a 4-sites plaquette in both lattices. In contrasted to the specially designed J-Q model in a square lattice \cite{JQ}, it may have sign problems. Earlier ED  in this model at $ Q/J \sim -1/2 $ in a triangular lattice points to a $ Z_2 $ QSL with the topological
degeneracy in a torus  \cite{frusrev}.

{\bf 6. Comments on the relations
 between phenomenological theory and microscopic calculations }

  In theoretical physics, there are two different and complementary approaches: one is  symmetry based effective  approach which may be used
  classify all the possible quantum phases and phase transitions. The advantage of this approach is that it may be able to list
  the possible phases. But the limitation is that which phase appears as  the  ground state of a given system can not be determined.
  Another approach is to perform a controlled calculation on a microscopic Hamiltonian.
  The advantage of this approach is it maybe able to identify the ground state and evaluate the excitations when there is a small parameter
  in the Hamiltonian. But its limitation is that no such controlled approach exists in the absence of such a small parameter.
  Then, powerful numerical calculations such as quantum Monte-Carlo simulations on a specific microscopic Hamiltonian
  may be performed to overcome such a limitation in some limited class of problems without sign problems.
  However, QMC may also suffer the notorious sign problem in most of the cases.
  In this work, we construct a  symmetry based effective approach which can be used to reproduce all the microscopic
  calculations at weak and strong couplings
  in \cite{gold}, then can be applied to predict the $ Z_2 $ QSL in the intermediate couplings.
    However, this combination of both approaches in a specific Hamiltonian may not be always possible. For example,
    in the dual vortex method constructed by the magnetic space group \cite{pq1,yan}, due to the highly non-local
    duality transformation from the boson to the vortex degree of freedoms, there is no way to establish the connections between
    the phenomenological parameters in the dual vortex theory and the bare parameters in any microscopic Hamiltonian.
    In the absence of the sign problem, it can still be compared with the QMC on a specific Hamiltonian.
    In \cite{yan}, despite not being able to establishing any such phenomenological/micrsoscopic  relations,
    the authors are still able to characterize the symmetry breaking patterns of the original bosons in the direct lattice by
    the gauge invariant density, kinetic energy and the currents of the dual vortices in the corresponding dual lattice.
    But this characterization is still symmetry based without establishing explicitly phenomenological/micrsoscopic connections.

Recently, there have been extensive research activities on the classifications of
topological phase of matter which break no symmetries \cite{tenfold,wenrev,senthil,xu,max}.
These phases also split into two classes:
interacting symmetry protected topological (SPT) phases
with trivial bulk order (short-range entanglement)
and symmetry enriched topological (SET) phases
with non-trivial  bulk topological order (long-range entanglement) \cite{tenfold,wenrev,senthil,xu,max}.
There could be also intrinsic connections between SPT and SET, for example, a 2d SET can act as
the surface states of a 3d SPT with boundary anomalies. Gauging SPT may lead to some classes of SET\cite{tenfold,wenrev}.
In some special cases, the Hamiltonian whose exact ground states
show specially designed SPT or SET orders can be constructed, but these Hamiltonians,
in general, involve highly non-local interactions which are needed to stabilize such states.
In most cases, the Hamiltonians which may host these phases are not known,
the classifications are purely symmetry based.
For a general simple experimental accessible Hamiltonian, microscopic analytical or numerical calculations
usually show  these states may have much higher energy than conventional symmetry broken states.
It remains  challenging to discover simple microscopic Hamiltonian on which one can perform microscopic calculations
to find any of these SPT or SET states. Of course, the $ Z_2 $ QSL is one of the simplest SET phase.
Here we provide a specific, simple and experimentally accessible system which hosts such a simple SET phase.

{\bf 7. The implications on cold atom and photonic experiments.-}

    In various promising material candidates, either geometrically frustrated systems or Kitaev materials,
    to search for possible QSLs \cite{SLrev1,SLrev3},
    there are always un-wanted  and un-controllable interactions which may un-stabilize the putative QSL in
    the theoretically designed Hamiltonians. Furthermore, due to the ubiquitous quenched disorders in materials,
    the quantum spin glass (QSG) \cite{SY,QSG1,QSG2} always competes seriously against the QSL. Both share similar properties
    except the topological properties. The cold atom systems provide un-precedent clean and tunable systems
    which can avoid these common difficulties suffered in these materials. Unfortunately, due to its diluteness,
    the cold atom system in optical lattices suffers its own difficulties: the heating problem.
    It is still under good control at the weak coupling regimes investigated in \cite{gold}, but will get worse as the interaction increases.
    Fortunately, as shown in Fig.\ref{QSL}, the QSL exists  at intermediate coupling strengths
    $ (U/t)_{c1} < U/t \sim 1 < (U/t)_{c2} $ where the heating issues  may still be controllable
    at the current stage of cold atom experiments. It may also be accessible to the photonic experiments.

    The salient features of a $ Z_2 $ QSL is its fractionized excitations with topological orders and long-range entanglements.
    One of the well known signatures of the deconfined spinons is its broad spin excitation spectrum even at $ T=0 $ in contrast to the sharp spin excitation spectrum in a magnetic ordered state. This feature has been observed in these materials at the lowest accessible
    temperatures by  in-elastic neutron scattering or resonant X-ray scattering techniques.
    In cold atoms,  it could also be easily detected by dynamic or elastic, energy or momentum resolved, longitudinal or
    transverse Bragg spectroscopies \cite{braggbog,braggangle,braggeng,braggsingle,becbragg,bragg12} in cold atoms
    and the site- and time-resolved spectroscopy in photonic systems \cite{gold}.
    Unfortunately, the QSG may also lead to similar behaviours.
    To settle down the issue, one must resort the more intrinsic and fundamental
    measurements on topological orders such as detecting the topological entanglement entropy to distinguish QSL from the QSG or other non-topological phases.
    This kind of smoking gun  measurements so far is quite difficult to carry out in materials \cite{tunneling}, but
    possible in the cold atom systems \cite{renyi}.
    In view of the more recent achievement on measuring the spin-charge de-confinement in a
    1d Fermionic Hubbard model \cite{bloch1}, $ Z_2 $ gauge theory in an optical lattice \cite{bloch2}
    and direct observation of incommensurate magnetism \cite{bloch3},
    if the heating issue can indeed be overcame,
    it is practical to directly probe the long-range entanglement  in the 2d QSL in Fig.\ref{QSL}.

{\bf 8. Conclusions and Perspectives. }

   As stressed in \cite{gold}, an Abelian flux in a bipartite lattice provides a new frustrating source than
   the geometric frustrations or Kitaev materials. Here we show that interacting spinor atoms
   moving in a square lattice subject to a $ \pi $ flux provides a new class of clean and tunable systems
   to search for still elusive QSL whose topological orders  maybe uniquely probed by current available cold atom experimental techniques.
   We also establish a deep and intrinsic
    connection between the Ginsburg-Landau effective action constructed here to describe the phases and transitions at any couplings  and
    the effective potential generated by the OFQD phenomena at weak coupling presented in \cite{gold}.
    This provides a new physical interpretation of the OFQD and enriches its impacts and applications considerably.
    The new method developed in this work can be transformed to study
    any quantum frustrated bosonic or fermionic systems in intermediate couplings.
    The  QSLs  may also appear  in these quantum frustrated fermionic systems in intermediate couplings.

 In Fig.\ref{QSL}, $ \mathbf{z}_1, \mathbf{z}_2 $ are the elementary excitations in the frustrated superfluids in the weak coupling limit.
While one of them is projected out, the other is confined into $ \vec{n} = \mathbf{z}^{\dagger} \vec{\sigma} \mathbf{z} $ in
the FM Mott state in the strong coupling limit.
However, in the $ Z_2 $ QSL sandwiched between the two states, there are gapped fractionalized
spinons $ z_{\alpha}, \alpha=1,2 $ which are dramatically different than $ \mathbf{z}_1, \mathbf{z}_2 $.
It was generally believed that the melting of a co-planar spin states leads to a QSL,
while that of a non-coplanar state leads to a chiral QSL \cite{CSL}.
Because Eq.\ref{piflux} in the $ \pi $ flux has the spin $ SU(2)_s $ symmetry, also Time reversal symmetry,
   so it should have no sign problem. Large scale QMC with soft core bosons or other numerical tools such as DMRG or Tensor networks
   maybe employed to investigate the nature of the $ Z_2 $ QSL \cite{softQMC} such as its topological entanglement entropy.
   Note that hard core bosons which can be mapped to quantum spin systems
   are more easily simulated by QMC, but not practical in cold atom experiments.
   It also satisfies the condition of Lieb-Schultz-Mattis (LSM) theorem: spin $ SU(2) $ symmetry and Time-reversal, spin $1/2 $ per unit cell,
   so any state which breaks no symmetries of the Hamiltonian must be a topological state with a long-range entanglement.

 It remains interesting to explore the nature of the two QCPs in Fig.\ref{QSL} from both the left and right.
 From the left frustrated SF side, it may be convenient to introduce a spin-anisotropic interaction
 $ V_{\lambda} =(1-\lambda) \sum_i S^2_{iz} $ to
 reduce the symmetry of the effective action Eq.\ref{general} from  $ SU(2)_s \times U(1)_c $ to $ U(1)_s \times U(1)_c $.
 In the easy-plane limit $ \lambda < 1 $, following the charge-vortex duality transformation in the
 bilayer quantum Hall system \cite{blqhye} which also has the  $ U(1)_s \times U(1)_c $ symmetry
 at any finite distance  $ d $ between the two layers, one can perform a charge-vortex duality transformation to study the frustrated SF to
 the QSL transition at $ (U/t)_{c1} $ in Fig.\ref{QSL}.
 Then some kind of paired vortices condensations may lead to the $ Z_2 $ QSL.
 From the right FM Mott side,
 the ring exchange terms discussed in Sec.5 were known to be important to drive a Mott state into a QSL \cite{SLrev1,SLrev3}.
 We will see how to construct a $ Z_2 $ gauge theory from the ring exchange terms.
 From the dual perspective, one may also need to see how to condense the fractionalized excitations such as $ e=z_a $ or $ m $ particle \cite{myown,fermion} to reach the  and the FM Mott on the right.
 We expect that (1) at $ (U/t)_{c1} $, condensing the $ e=z_a $ particle leads to the frustrated SF on the left,
 (2) at $ (U/t)_{c2} $, condensing the $ m $ particle of the $ Z_2 $ gauge fields lead to the confinement of $ z_a $ into $ \vec{n}=\mathbf{z}^{\dagger} \vec{\sigma} \mathbf{z} $, therefore reach the FM Mott on the right ( Fig.1).
 The effects of a Zeeman field will also be investigated in a separate publication.

{\bf Acknowledgements }

 J. Ye thank Gang Chen and Jiansheng Wu for interesting discussions. We acknowledge AFOSR FA9550-16-1-0412 for supports.

{\bf Appendix }

 In this appendix, we
(1) establish the nature of the $ Z_2$ QSL.
(2) comment on the relation between the GL effective action and WZW model with or without  a topological term
(3) Apply our method to the simpler case of single component boson subject to a $ \pi $ flux and show the absence of OFQD
    phenomena at weak coupling.

{\bf A. Contrast the Spin structure in a square lattice with that in a triangular lattice and a possible $ Z_2 $ QSL }

The spin structure of a general Mott state in the main text can be simplified when confined to the
classical mean-field ground-state
$\mathbf{z}_1^\dagger\mathbf{z}_2=0$. Then can be simplified further if imposing the constraint from
the OFQD.

First, a simple algebra shows
\begin{align}
    (\mathbf{z}_1^\dagger\sigma^a \mathbf{z}_1)(\mathbf{z}_1^\dagger\sigma^a \mathbf{z}_1)
	=(\mathbf{z}_1^\dagger\mathbf{z}_1)^2
\end{align}
and similar relation
$(\mathbf{z}_2^\dagger\sigma^a \mathbf{z}_2)(\mathbf{z}_2^\dagger\sigma^a \mathbf{z}_2)
=(\mathbf{z}_2^\dagger\mathbf{z}_2)^2$
holds.
In addition, we can also obtain
\begin{align}
    (\mathbf{z}_1^\dagger\sigma^a \mathbf{z}_1)(\mathbf{z}_2^\dagger\sigma^a \mathbf{z}_2)
	&=2(\mathbf{z}_1^\dagger\mathbf{z}_2)(\mathbf{z}_2^\dagger\mathbf{z}_1)
	-(\mathbf{z}_1^\dagger\mathbf{z}_1)(\mathbf{z}_2^\dagger\mathbf{z}_2)    \nonumber  \\
    (\mathbf{z}_1^\dagger\sigma^a \mathbf{z}_2)(\mathbf{z}_1^\dagger\sigma^a \mathbf{z}_2)
	&=(\mathbf{z}_1^\dagger\mathbf{z}_2)^2    \nonumber  \\
    (\mathbf{z}_1^\dagger\sigma^a \mathbf{z}_2)(\mathbf{z}_2^\dagger\sigma^a \mathbf{z}_1)
	&=2(\mathbf{z}_1^\dagger\mathbf{z}_1)(\mathbf{z}_2^\dagger\mathbf{z}_2)
	-(\mathbf{z}_1^\dagger\mathbf{z}_2)(\mathbf{z}_2^\dagger\mathbf{z}_1)
\end{align}
and more generally we have
$(\mathbf{z}_i^\dagger\sigma^a \mathbf{z}_j)(\mathbf{z}_k^\dagger\sigma^a \mathbf{z}_l)
=2(\mathbf{z}_i^\dagger\mathbf{z}_l)(\mathbf{z}_k^\dagger\mathbf{z}_j)
-(\mathbf{z}_i^\dagger\mathbf{z}_j)(\mathbf{z}_k^\dagger\mathbf{z}_l)$.
Then it is easy to verify the norm of $\vec{n}_{1,2,3}$
\begin{align}
    (\vec{n}_1)^2
	&=(\mathbf{z}_1^\dagger\mathbf{z}_1-\mathbf{z}_2^\dagger\mathbf{z}_2)^2
	+4(\mathbf{z}_1^\dagger\mathbf{z}_2)(\mathbf{z}_2^\dagger\mathbf{z}_1)  \nonumber \\
    (\vec{n}_2)^2
	&=(\mathbf{z}_1^\dagger\mathbf{z}_2-\mathbf{z}_2^\dagger\mathbf{z}_1)^2
	+4(\mathbf{z}_1^\dagger\mathbf{z}_1)(\mathbf{z}_2^\dagger\mathbf{z}_2)  \nonumber \\
    (\vec{n}_3)^2
	&=-(\mathbf{z}_1^\dagger\mathbf{z}_2+\mathbf{z}_2^\dagger\mathbf{z}_1)^2
	+4(\mathbf{z}_1^\dagger\mathbf{z}_1)(\mathbf{z}_2^\dagger\mathbf{z}_2)
\end{align}
and the dot products between  $\vec{n}_{1,2,3}$
\begin{align}
    \vec{n}_1\cdot\vec{n}_2
	&=(\mathbf{z}_1^\dagger\mathbf{z}_2+\mathbf{z}_2^\dagger\mathbf{z}_1)
	(\mathbf{z}_1^\dagger\mathbf{z}_1+\mathbf{z}_2^\dagger\mathbf{z}_2)   \nonumber \\
    \vec{n}_1\cdot\vec{n}_3
	&=i(\mathbf{z}_1^\dagger\mathbf{z}_2-\mathbf{z}_2^\dagger\mathbf{z}_1)
	(\mathbf{z}_1^\dagger\mathbf{z}_1+\mathbf{z}_2^\dagger\mathbf{z}_2) \nonumber \\
    \vec{n}_2\cdot\vec{n}_3
	&=i[(\mathbf{z}_1^\dagger\mathbf{z}_2)(\mathbf{z}_1^\dagger\mathbf{z}_2)
	    -(\mathbf{z}_2^\dagger\mathbf{z}_1)(\mathbf{z}_2^\dagger\mathbf{z}_1)]
\end{align}

For the density distribution, one can replace $\sigma$ by 1 and obtain
\begin{align}
	n_i &=\mathbf{z}_n^\dagger \mathbf{z}_n + \frac{1}{\sqrt{2}}
	[e^{i\pi/4}\mathbf{z}_1^\dagger \mathbf{z}_2
	+e^{-i\pi/4}\mathbf{z}_2^\dagger\mathbf{z}_1](-1)^{i_x}  \nonumber  \\
	& +\frac{1}{\sqrt{2}}
	[e^{-i\pi/4}\mathbf{z}_1^\dagger \mathbf{z}_2
	+e^{i\pi/4}\mathbf{z}_2^\dagger\mathbf{z}_1](-1)^{i_y}
\end{align}
which may also expressed as
\begin{align}
	n_i &=\mathbf{z}_n^\dagger \mathbf{z}_n
	   +\frac{1}{2}[(\mathbf{z}_m^\dagger \tau_{mn}^x\mathbf{z}_n)
	   		+(\mathbf{z}_m^\dagger \tau_{mn}^y\mathbf{z}_n)](-1)^{i_x}   \nonumber  \\
	   & +\frac{1}{2}[(\mathbf{z}_m^\dagger \tau_{mn}^x\mathbf{z}_n)
	   		-(\mathbf{z}_m^\dagger \tau_{mn}^y\mathbf{z}_n)](-1)^{i_y}
\end{align}

It is easy to verify that $\sum_i n_i^2$ lead to the $u_1=2u_2=u $ term
in the Lagrangian Eq.\ref{general}.

$\bullet$
If we impose the mean-field condition $\mathbf{z}_1^\dagger\mathbf{z}_2=0$ which holds in the $ U \rightarrow \infty $ limit,
then $ n_i=\mathbf{z}_n^\dagger \mathbf{z}_n =n $.  It also leads to a great simplification in the spin sector:
\begin{align}
	& \vec{n}_1^2 =(\mathbf{z}_1^\dagger\mathbf{z}_1-\mathbf{z}_2^\dagger\mathbf{z}_2)^2,\quad
	\vec{n}_2^2=\vec{n}_3^2=4(\mathbf{z}_1^\dagger \mathbf{z}_1)(\mathbf{z}_2^\dagger \mathbf{z}_2),\nonumber  \\
	& \vec{n}_1\cdot\vec{n}_2=\vec{n}_1\cdot\vec{n}_3=\vec{n}_2\cdot\vec{n}_3=0
\end{align}
 which also implies the relation $(\vec{n}_2+\vec{n}_3)\cdot(\vec{n}_2-\vec{n}_3)=0$.

So the mean-field spin has the ``length''
\begin{align}
	\vec{S}_i\cdot\vec{S}_i
	=\vec{n}_1^2+\frac{1}{2}(\vec{n}_2^2+\vec{n}_3^2)
	=(\mathbf{z}_1^\dagger\mathbf{z}_1+\mathbf{z}_2^\dagger\mathbf{z}_2)^2
	=n^2
\end{align}
 So the density is just $n_i=n=\sqrt{\vec{S}_i\cdot\vec{S}_i}$.

$\bullet$
In addition to imposing the condition $\mathbf{z}_1^\dagger\mathbf{z}_2=0$, one also
impose the OFQD condition at the easy plane limit
 $\mathbf{z}_1^\dagger\mathbf{z}_1=\mathbf{z}_2^\dagger\mathbf{z}_2=n/2$ which holds in the
 $ w \rightarrow \infty $ limit, then it leads to $ \vec{n}_1=0 $ and
\begin{align}
	 \vec{n}_2^2=\vec{n}_3^2=n^2, ~~~\vec{n}_2\cdot\vec{n}_3=0
\end{align}
and the spin is coplanar
\begin{align}
	\vec{S}_i=\frac{1}{2}(\vec{n}_2+\vec{n}_3)(-1)^{i_x}+\frac{1}{2}(\vec{n}_2-\vec{n}_3)(-1)^{i_y}
\label{twoorderingixiy}
\end{align}
  with a uniform density  $n_i=n$.

  So we show that in the hard core limit $ U \rightarrow \infty $ and strong easy-plane limit $ w \rightarrow \infty $,
  the low energy physics can be expressed in terms of the quantum fluctuations of the two orthogonal unit vectors $ \vec{n}_2 $ and
  $ \vec{n}_3 $. The spins can be expressed in terms of $ \vec{n}_2 $ and $\vec{n}_3 $ in Eq.\ref{twoorderingixiy} with only
  two ordering wavevectors $ (\pi,0) $ and $ (0,\pi) $.

This can be contrasted to the 3-sublattice  $ 120^{\circ} $ coplanar
spin structure Mott state in a triangular lattice\cite{frusrev,sachdev} which can be written as:
\begin{equation}
 \vec{S}= \vec{n}_1 \cos (\vec{Q} \cdot \vec{x} ) + \vec{n}_2 \sin (\vec{Q} \cdot \vec{x} )
\label{onetri}
\end{equation}
 where   $ \vec{Q}= \frac{4 \pi}{a} (1/3,1/\sqrt{3} ) $ is the only  ordering wavevector.
The two orthogonal unit vectors  $ \vec{n}_1, \vec{n}_2 $
satisfying $ \vec{n}^2_1= \vec{n}^2_2=1,  \vec{n}_1 \cdot \vec{n}_2=0 $
can be written in terms of two spinons  as $ n_{2 \alpha} + i n_{1 \alpha} = \epsilon_{ac} z_c \sigma^{\alpha}_{ab} z_b $
where $ |z_1|^2 + |z_2|^2=1 $. For the quantum anti-ferromagnetic Heisenberg model in a triangular lattice,
the spinons condense $ \langle z_\alpha \rangle \neq 0 $, so leads to the  3 sublattice $ 120^{\circ} $ coplanar spin structure.
If under some perturbations such as the next nearest neighbor coupling $ J_2 $,
they may not condense, then it becomes a $ Z_2 $ QSL with gapped deconfined bosonic spinons $ z_{\alpha} $.
The $ Z_2 $ gauge field does not show up in the continuum limit, but plays important roles in the topological orders of
this QSL. Here one need only replace the  $ \vec{n}_1, \vec{n}_2 $ and Eq.\ref{onetri} in the triangular lattice by $   \vec{n}_2, \vec{n}_3 $
and Eqn.\ref{twoorderingixiy} in the square lattice, then the deconfined spinons are also $ z_{\alpha}, \alpha=1,2 $
carrying spin $ s=1/2 $.

 In reality, the onsite Hubbard interaction $ U $ is finite, in fact, it is in the intermediate range
 $ (U/t)_{c1} < U/t < (U/t)_{c2}  $ in Fig.\ref{QSL}, so indeed there are large charge fluctuations.
 The $ w $ term which is  generated by the OFQD changes is very samll
 in the weak coupling limit, but may increase as the interaction increases as shown in Fig.\ref{QSL}.
 Although one has no analytical evaluation of its value at intermediate couplings, it must change sign  at $ (U/t)_{c2} $,
 so its value become small also near the QSL to the Mott FM transition.
 Note that  the 3 sublattice $ 120^{\circ} $ coplanar
spin state in a triangular lattice can be achieved just at the classical level \cite{frustri}, while the  4 sublattice $ 90^{\circ} $ coplanar
spin state in a square lattice  is selected by quantum fluctuations only. This fact suggests that the latter is more vulnerable
to melting into the $ Z_2 $ QSL than the former.
Therefore, we argue that these two sources of quantum fluctuations will melt the
4- sublattice $ 90^{\circ} $ coplanar spin structure into the $ Z_2 $ QSL.

{\bf B. Constructing the order parameters in terms of a $U(2) $ matrix
and possible connections to $2+1 $ dimensional WZW model with or without a topological term. }

Very intuitively, one can introduce a $U$ matrix as
\begin{align}
	U=(\mathbf{z}_1,\mathbf{z}_2)
	=\begin{pmatrix}
		z_{1\uparrow} &z_{2\uparrow}\\
		z_{1\downarrow} &z_{2\downarrow}\\
	\end{pmatrix}
	=\begin{pmatrix}
		z_{11} &z_{21}\\
		z_{12} &z_{22}\\
	\end{pmatrix}
\label{U2}
\end{align}
then it is easy to verify
\begin{align}
	U^\dagger U=
	\begin{pmatrix}
		z_{1\uparrow}^*z_{1\uparrow}+z_{1\downarrow}^*z_{1\downarrow}
			&z_{1\uparrow}^*z_{2\uparrow}+z_{1\downarrow}^*z_{2\downarrow}\\
		z_{2\uparrow}^*z_{1\uparrow}+z_{2\downarrow}^*z_{1\downarrow}
			&z_{2\uparrow}^*z_{2\uparrow}+z_{2\downarrow}^*z_{2\downarrow}\\
	\end{pmatrix}
\end{align}

One can also decompose them as:
\begin{align}
	U^\dagger U
	&=\frac{1}{2}(\mathbf{z}_1^\dagger\mathbf{z}_1+\mathbf{z}_2^\dagger\mathbf{z}_2)\sigma_0
	+\frac{1}{2}(\mathbf{z}_1^\dagger\mathbf{z}_2+\mathbf{z}_2^\dagger\mathbf{z}_1)\sigma_x    \nonumber  \\
	&+\frac{i}{2}(\mathbf{z}_1^\dagger\mathbf{z}_2-\mathbf{z}_2^\dagger\mathbf{z}_1)\sigma_y
	+\frac{1}{2}(\mathbf{z}_1^\dagger\mathbf{z}_1-\mathbf{z}_2^\dagger\mathbf{z}_2)\sigma_z
\label{udu}
\end{align}
  where $\sigma_0 = I $ denotes the 2 by 2 identity matrix.

If one need to impose both the orthogonal condition $\mathbf{z}_1^\dagger\mathbf{z}_2=0$
and the equal magnitude condition $\mathbf{z}_1^\dagger\mathbf{z}_1=\mathbf{z}_2^\dagger\mathbf{z}_2=n/2$,
then $U^\dagger U=(n/2)\sigma_0$.
If one rescales $U$ by $\sqrt{2/n}$ to make $U^\dagger U=\sigma_0$,
then $UU^\dagger=\sigma_0$.

When taking the saddle point solution of $U$ as a U(2) matrix,
then from Eq.\ref{udu}, one can see that the fluctuation {\em out of } the classic degenerate manifold is off-diagonal in terms of $U^\dagger U$,
while the OFQD fluctuation {\em within } the  classic degenerate manifold
is diagonal and $\propto \sigma_z$.

In the $ U \rightarrow \infty $ and $ w\rightarrow \infty $ limit,
The effective action in the intermediate couplings $ (U/t)_{c1} < U/t < (U/t)_{c2}  $ can be described by the
$ 2+1 $ dimensional  WZW model with a possible topological term:
\begin{align}
  & {\cal L}_{WZW} = \frac{1}{2 g } Tr( \partial_{\mu} U^{\dagger}  \partial_{\mu} U ) + i \theta Q,   \nonumber   \\
  & Q  = \frac{1}{ 24 \pi^2} \epsilon_{\mu \nu \lambda} Tr[ U^{\dagger} \partial_{\mu} U
   U^{\dagger} \partial_{\nu} U U^{\dagger} \partial_{\lambda} U ]
\label{wzw}
\end{align}
  where $ \mu, \nu, \lambda=\tau, x, y $ are space-time dimension,
  $ U $ is a $ U(2) $ matrix Eq.\ref{U2},
  $ Q $ is the integer winding number due to $ \Pi_{3} ( S^3 ) = Z $.
  Our microscopic calculation seems suggest the absence of topological term, so $ \theta=0 $.
  It can be shown that due to $ U(2)=SU_s(2) \times U_c(1) $, the charge sector $ U(1)_c $ decouples from the
  spin $ SU_s(2) $ sector.
  When $ g < g_c $, $ \langle U \rangle \neq 0 $ breaks the  $  SU(2)_s $ symmetry to $ 1 $ leads to
  the 4-sublattice $90^{\circ} $ co-planar Mott phase and the 3 linear gapless modes above the ground state.
  However, as argued in the last section, the strong quantum fluctuations due to the finite intermediate $ U $
  and small $ w $ not included in Eq.\ref{wzw} will likely melt this Mott state into a $ Z_2 $ QSL shown in Fig.\ref{QSL}.

  However, $ \theta =\pi $ was argued in \cite{senthil,xu,max} to described the surface state of a symmetry protected 3d
  bosonic topological state (SPT) with the axion angle $ \Theta=2 \pi $ in
  $ \frac{\Theta}{ 4 \pi^2} \vec{E} \cdot \vec{B} $. If so, it may also lead to an exotic $ Z_2 $ QSL phase
  where both $ e $ and $ m $ particles carry spin $ s=1/2 $. This exotic $ Z_2 $ QSL is different than the $ Z_2 $ QSL
  in a square lattice in Fig.\ref{QSL}, so it can only be realized as a surface state of a 3d SPT phase.

  In fact, as shown in \cite{senthil,xu}, Eq.\ref{wzw} can be written in terms of more intuitive
  $ O(4) $ Non-linear $ \sigma $ model.  By writing $ U= n_0 + i \vec{n} \cdot \vec{\sigma} $
  subject to the constraint $ n^2_0+ (\vec{n} )^2=1 $ and defining $ \vec{N}= ( n_0, \vec{n} ) $ which
  is a 4-component unit vector in $ S^4 $,  then  Eq.\ref{wzw} can be -rewritten as a $ SO(4) $ non-linear sigma model (NLSM):
\begin{align}
   &{\cal L}_{O(4)}  =  \frac{1}{2 g } ( \partial_{\mu} \vec{N}  )^2 + i \theta Q,  \nonumber   \\
   & Q = \frac{1}{ 12 \pi^2} \epsilon_{\mu \nu \lambda} \epsilon_{\alpha \beta \gamma \delta}
   N_{\alpha} \partial_{\mu} N_{\beta} \partial_{\nu} N_{\gamma} \partial_{\lambda} N_{\delta}
\label{o4}
\end{align}
   where $ \mu, \nu, \lambda=\tau, x, y $ are space-time dimension, $ \alpha, \beta, \gamma, \delta $  are
   the four components. Then in this case $ \theta=\pi $ corresponds to the deconfined QCP between the
   Neel state denoted by $ \vec{n} $ and a Valence bond solid state denoted by $ n_0 $.
   As shown in \cite{senthil}, Eq.\ref{o4} can also be extended to $ 3+1 $ d $ SO(5) $ group with $ \vec{N} $ a five-component unit vector and the topological charge $ Q $ due to $ \Pi_4 ( S^4)= Z $. This $ SO(5) $ NLSM with $ \theta=\pi $ may be used to describe
   the bulk of the 3d bosonic SPT.

{\bf C. One component boson in $ \pi $ flux }

 In the main text, we studied two component bosons in a $ \pi $ flux. In Sec.4, we also studied the  two component bosons without
 a flux ( See Eq.\ref{singlecomponent} ).
 Here, we will study one component boson in $ \pi $ flux. The method developed in main text can be easily applied to this much simpler case also.
 In this case, we focus also on the $ n=1 $ filling, so the filling factor is also $ \nu=2 $. We did not see the BIQH claimed in \cite{senthil}
 in this simpler case either. The main purpose here is to show there is no OFQD in the one component case.

  Denoting the square lattice site by $ \vec{x} = (a_1, a_2 ) $.
  In the Landau gauge used in \cite{yan}, the boson operator can be written as:
\begin{eqnarray}
   \psi( \vec{x} ) & = &  \sum^{1}_{m=0}  c_{m} e^{ i \pi m a_1 }[  \xi_0
   + \xi_1 \omega^{-m} e^{ i \pi a_2 } ]     \nonumber  \\
   & = & \sum^{1}_{m=0}  c_{m} (-1)^{ m a_1 }[  \xi_0 + \xi_1 (-1)^{m} (-1)^{ a_2 } ]
\label{onecompfield}
\end{eqnarray}
   where $ c_m=(1, \sqrt{2}-1) $

   In the permutative representation:
\begin{eqnarray}
  \xi_0 & = &  ( \phi_0 + \phi_1 )/\sqrt{2}   \nonumber  \\
  \xi_1 & = &  -i( \phi_0 - \phi_1 )/\sqrt{2}
\end{eqnarray}

 It is also easy to see the boson density
 $ n= \psi^{\dagger} ( \vec{x} ) \psi( \vec{x} )=|\xi_0|^2+  |\xi_1|^2  = |\phi_0|^2+  |\phi_1|^2 $.

By diagonizing the boson kinetic term directly in the gauge chosen in Fig1. of Ref.\cite{gold}, we find
the permutative representation directly. So the advantage of the gauge chosen in Fig1. of Ref.\cite{gold}
over the Landau gauge used in \cite{pq1,yan} is that it directly leads to  permutative representation.
    Upto 8th order, one can show that the effective action, consist with all the
    Time reversal $ {\cal T} $ and the $ U(1)_c $, the
    MSG consisting of the two translations $ T_x, T_y $, rotation $ R_{\pi/2} $, the two reflections
    $ I_x, I_y $, can be written as  $ {\cal L}_{SF}   ={\cal L}_{0} + {\cal L}_{1} + {\cal L}_{2} $:
\begin{eqnarray}
    {\cal L}_{0} & = &  \phi_{l}^{\dagger} \partial_{\tau} \phi_l + v^2 |  \nabla \phi_{l} |^{2} + r | \phi_{l} |^{2} + u ( | \phi_{l} |^{2} )^2 + \cdots     \nonumber  \\
    {\cal L}_{1} & =  &  w  ( | \phi_{0} |^{2} - |\phi_{1} |^{2} )^{2}
                     \nonumber  \\
     {\cal L}_{2} & = &  \lambda [ (\phi^{*}_{0} \phi_{1} )^{4} +h.c.]
\label{ka}
\end{eqnarray}
     where $ v^2, r, u $  and $ w,  \lambda $ are phenomenological parameters.
     Unfortunately, the signs of $ w, \lambda $ can only be determined by microscopic calculations.

   In the weak coupling limit $ U/t \ll 1 $, it is in the SF side where $ r < 0 $.
   In the Ising case $ w < 0 $:  $  \phi_0=1,  \phi_1=0 $ or vice versa.
   So the state has the degeneracy $ 2 $ corresponding to which of the 2 fields condenses.
   The boson fields, kinetic energies and currents are shown in Fig.19a in \cite{yan} which was
   used here. Due to its chiral boson currents, it can be called a chiral SF state.

\begin{figure}
\includegraphics[width=4cm]{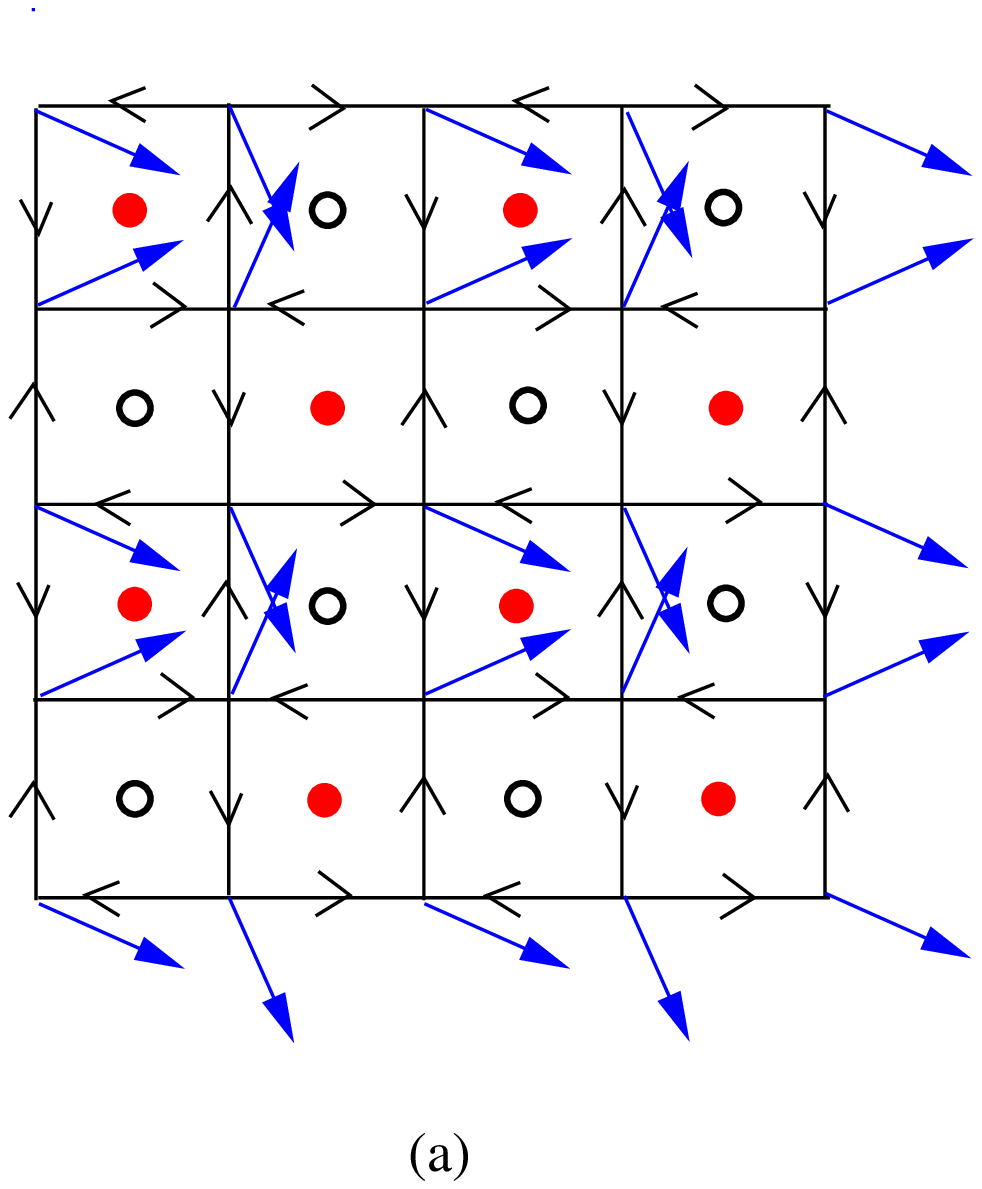}
\hspace{0.2cm}
\includegraphics[width=3.5cm]{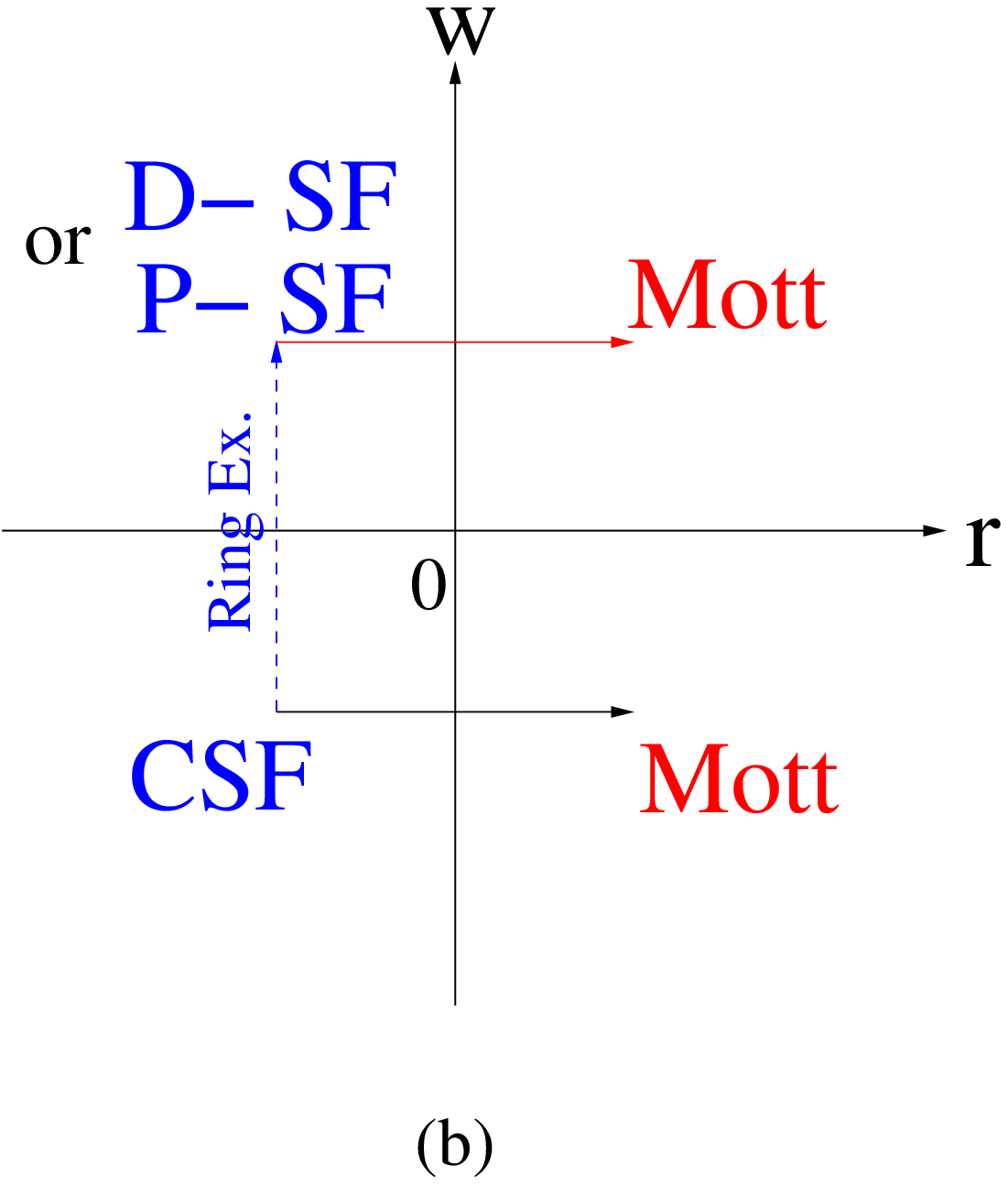}
\caption{ (a) The chiral SF state (CSF) in the Ising limit  $ w < 0 $ in  a square lattice.  It is 2 fold degenerate.
One of them is shown here. The other is just taking the c.c of it.
  The boson field $ \psi( 0,0) =   1-( \sqrt{2}-1)i, \psi( 0,1) =  1+ ( \sqrt{2}-1)i,
  \psi( 1,0) =   ( \sqrt{2}-1)- i, \psi( 1,1) =  ( \sqrt{2}-1) + i $.
  The boson current flowing counter clockwise is $ I=  \sqrt{2}-1 $. All the bonds have the same boson kinetic energy $ K= \sqrt{2}-1 $.
  There are no frustrated bonds. The red solid dot stands for a SF vortex, the empty dot an anti-vortex.
  It clearly breaks the Time-reversal symmetry. It is the true ground state with only on-site intercation.
  (b) The three possible phases of the one component bosons in the presence of $ \pi $ flux.
  D-SF means Dimer SF, P-SF means Plaquette SF.
  Independent of Ising or Easy-plane limit, it only has one Mott phase. Compare with Fig.1. }
\label{fig19}
\end{figure}

  In the easy plane limit \cite{pq1} $ w > 0 $, there are also two cases depending on the signs of the quartic term  $ \lambda \cos 4 \theta $:

  (A) If $ \lambda > 0 $, then $ \phi_0= \phi_1 e^{i n \pi/2 }, n=0,1,2,3 $.
      So the state has a degeneracy $ 4 $ corresponding to the 4 possible ways to condense the  2 boson fields.
      The boson fields, kinetic energies of the $ n=0 $ case with  $ \phi_0= \phi_1=1 $ is shown in Fig.20a of Ref.\cite{yan}.
      It is a dimer SF state.

  (B) If $ \lambda < 0 $, then $ \phi_0= \phi_1 e^{i ( n+1/2 ) \pi/2 }, n=0,1,2,3 $.
      So the state has a degeneracy $ 4 $ corresponding to the 4 possible ways to condense the  2 boson fields.
      The boson fields, kinetic energies of the $ n=0 $ case with  $ \phi_1=1 $ is shown in Fig.20b of Ref.\cite{yan}.
      It is a plaquette SF state.

\begin{figure}
\includegraphics[width=3cm]{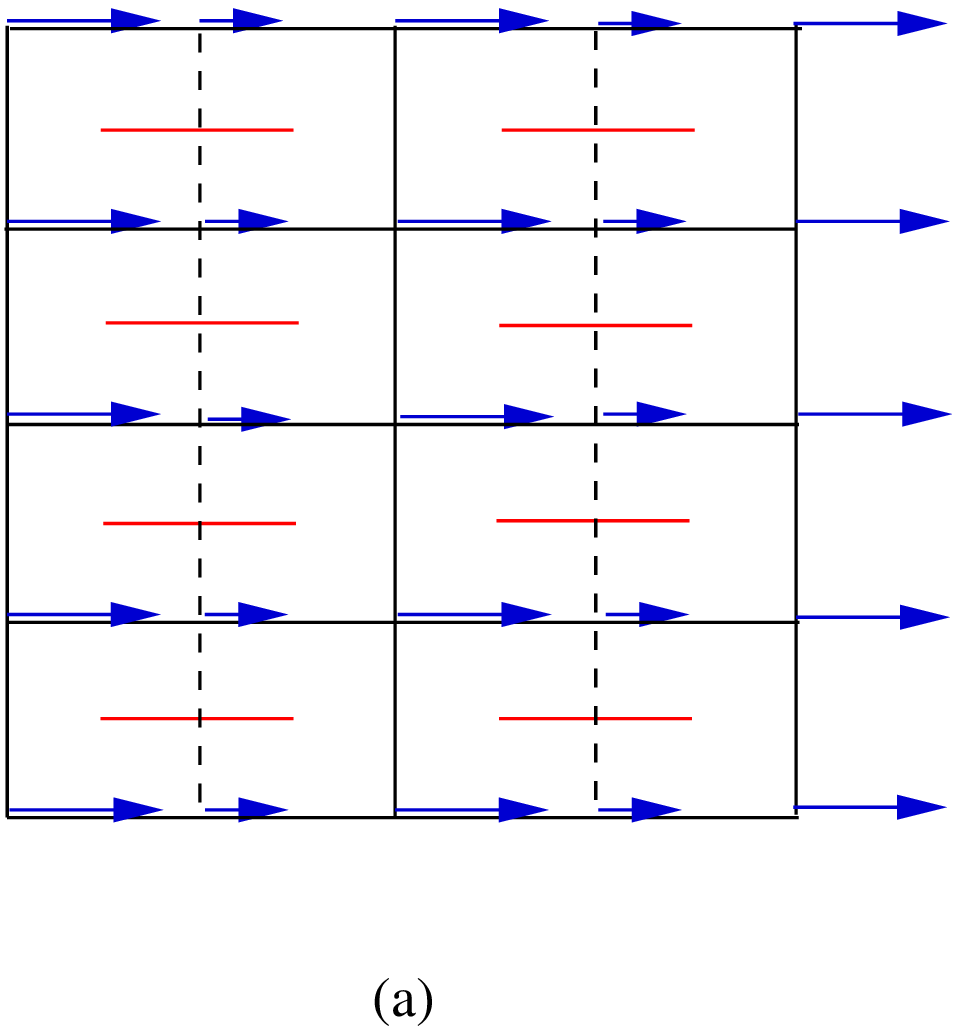}
\hspace{0.5cm}
\includegraphics[width=3cm]{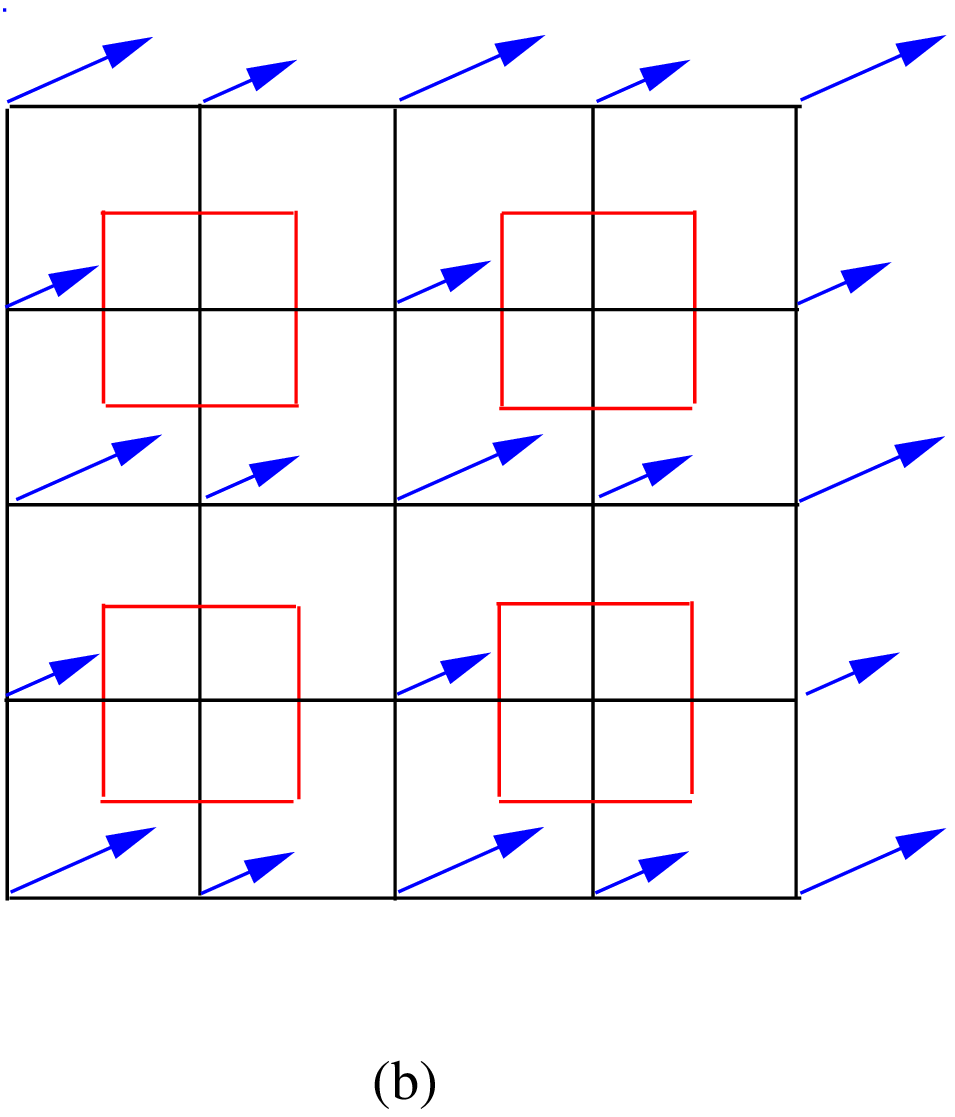}
\caption{ The ground states in the easy-plane limit $ w > 0 $ in a square lattice
(a) Dimer SF state at $ \lambda > 0 $. It is 4 fold degenerate. One of them is shown here.
The long arrow is $ \sqrt{2} $,
the short arrow is $ \sqrt{2} ( \sqrt{2}-1) $. The dashed line is the frustrated boson bond with the strength $ K_{fv}=-2 (\sqrt{2}-1)^{2} $.
The un-frustrated vertical vertex bond is $ K_{v}=2 $. The horizontal bond is $ K_{h}=  2 (\sqrt{2}-1) $.
The red bond which is perpendicular to the frustrated boson bond is the vortex bond in the dual lattice.
The vortices are hopping back and forth along the red bond.
(b) Plaquette SF state at $ \lambda < 0 $. It is also 4 fold degenerate. One of them is shown here.
The boson field  $ \psi( 0,0) = 2  [1+ ( \sqrt{2}-1)i], \psi( 0,1) = \sqrt{2} [1+ ( \sqrt{2}-1)i],
  \psi( 1,0) = \sqrt{2} [1+ ( \sqrt{2}-1)i]= \psi( 0,1), \psi( 1,1) = 0 $. The arrows indicate both the magnitude and the phase of the boson fields.  Because the boson field vanishes at $ (1,1) $, so there is a local vortex flowing
  around this lattice point as indicated by the red plaquette. The vortices are hopping around the red plaquette.
  The kinetic energies emanating from the zero boson field points are zero.  All the other bonds have strength $ K= 8 ( \sqrt{2}-1) $.
   There are no frustrated bonds.  There are no boson currents in both (a) and (b) indicating no
  time-reversal symmetry breaking ( or no chirality ). The easy-plane limit is not realized with only on-site interaction $ U $,
  but it still could be realized with adding some ring exchange terms.  }
\label{fig18}
\end{figure}

  At weak coupling limit $ U/t $, one can perform a weak coupling expansion.
  Plugging in Eq.\ref{onecompfield} into the Hamiltonian, one can directly find the expression of the
  phenomenological parameters $ r, u, v $  and $ w, \lambda $ in Eq.\ref{ka} in terms of the microscopic
  parameters.  At weak coupling, we find $ w=-4 U^2/t < 0  $ without involving the OFQD phenomena.
  It is indeed in the Ising limit where Fig.\ref{fig19} holds.
  So the easy-plane limit shown in Fig.\ref{fig18} is irrelevant with only on-site interaction at weak coupling.
  But it could become a ground state when adding some ring exchange interactions which could change $ w $ to positive as shown
  along the dashed vertical line in Fig.\ref{fig19}b.

  At strong coupling limit $ U/t \gg 1 $, it is a Mott state at the $ n=1 $ filling where $ r >0 $.
  Because the boson density $ n=  |\phi_0|^2+  |\phi_1|^2 $, Ising or Easy-plane limit leads to only one Mott phase.
  So we expect $ w < 0 $ extends to the strong coupling as shown along the black solid horizontal line in Fig.\ref{fig19}b.
  Therefore there is no intervening phase between the CSF at weak coupling and the Mott phase at the strong coupling.
  However, due to the Time-reversal symmetry breaking of the CSF, the universality class from the Mott to the CSF remains
  to be explored.
  This is in sharp contrast to the two component spinor boson case where the it changes from the easy-plane limit at weak coupling to
  the Ising limit in the strong coupling, so there must be some intermediate phases between the two limits.
  The comparison between Fig.2 and Fig.1 may inspire one to look at if adding NN $ V_1 $ or even NNN $ V_2 $ density-density interactions
  to Eq.1 will change the easy-plane limit $ w >0 $ to the Ising limit in Fig.1.


\begin{thebibliography}{99}
\bibitem{aue} A. Auerbach,
\textit{Interacting electrons and quantum magnetism},
(Springer Science \& Business Media, 1994).


\bibitem{sachdev} S. Sachdev,
\textit{Quantum Phase transitions},
(2nd edition, Cambridge University Press, 2011).

\bibitem{frusrev} For a review, see
C. Lhuillier and G. Misguich, Frustrated quantum magnets, arXiv:cond-mat/0109146


\bibitem{and1} P. W. Anderson, Resonating valence bonds: A new kind of
insulator. Mater. Res. Bull. 8, 153每160 (1973). doi: 10.1016/
0025-5408(73)90167-0

\bibitem{and2} P. W. Anderson, The resonating valence bond state in
La2CuO4 and superconductivity. Science 235, 1196每1198
(1987). doi: 10.1126/science.235.4793.1196; pmid: 17818979

\bibitem{subir} Subir Sachdev,
Kagome∩- and triangular-lattice Heisenberg antiferromagnets: Ordering from quantum fluctuations and quantum-disordered ground states with unconfined bosonic spinons, Phys. Rev. B 45, 12377 (1992) - Published 1 June 1992

\bibitem{dimer1}  Daniel S. Rokhsar and Steven A. Kivelson, Superconductivity and the Quantum Hard-Core Dimer Gas, Phys. Rev. Lett. 61, 2376 每 Published 13 November 1988.

\bibitem{dimer2}  R. Moessner and S. L. Sondhi, Resonating Valence Bond Phase in the Triangular Lattice Quantum Dimer Model,
Phys. Rev. Lett. 86, 1881 (2001).

\bibitem{dimer3} Hong Yao and Steven A. Kivelson,
Exact Spin Liquid Ground States of the Quantum Dimer Model on the Square and Honeycomb Lattices, Phys. Rev. Lett. 108, 247206 (2012).

\bibitem{largeN} N. Read, S. Sachdev, Large-N expansion for frustrated
quantum antiferromagnets. Phys. Rev. Lett. 66, 1773每1776
(1991). doi: 10.1103/PhysRevLett.66.1773; pmid: 10043303

\bibitem{largeN2} S. Sachdev and N. Read, Large-N expansion for frustrated
and doped quantum antiferromagnets, Int. J. Mod. Phys. B 5, 219 (1991).

\bibitem{wen1} X. G. Wen, Mean-field theory of spin-liquid states with finite energy gap and topological orders,
Phys. Rev. B 44, 2664 每 Published 1 August 1991

\bibitem{wen2} Xiao-Gang Wen, Quantum orders and symmetric spin liquids,  Phys. Rev. B 65, 165113 每 Published 10 April 2002

\bibitem{Kit1}  A. Yu. Kitaev, Fault-tolerant quantum computation by anyons. Ann. Phys. 303, 2每30 (2003).

\bibitem{myown} Jinwu Ye, Quantum generated vortices, dual singular gauge transformation and
       zero temperature transition from d-wave superconductor to underdoped
       regime.  Phys. Rev. B. 65, 214505 (2002).
In fact, historically, the $ Z_2 $ gauge theory and the long-range semonic mutual statistics first happens in
the ordinary type II S-wave superconductors: the fermionic Bogoliubov quasi-particle winding around the $ hc/2e $ vortex
 acquires a phase $ \pi $, so the  $ \epsilon $ particle and the $ hc/2e $ vortex are mutual semions.
There is a mutual $ Z_2 $ Chern-Simon term between the two.
This has been discussed in Sec.III of this work, in a type II S-wave superconductor, the quasi- particle has a gap,
Due to the Messiner effects, the vortices are only short-range interacting,
so the mutual statistical  interaction between the quasi- particle and the $ hc/2e $ vortex is the only long-range interaction
in the system leads to the long-range entaglement in the system. So the the fermionic Bogoliubov quasi-particleand
 the $ hc/2e $ vortex can be mapped to the $ \epsilon $ and $ m $ particle in the  $ Z_2 $ QSL ( or 2d/3d Toric code ).
 The confinement transition driven by the condensation of the  $ hc/2e $ vortices was also discussed.
Unfortunately, using a $ U(1) $ CS term in a continuous system as did in this work to represent this mutual statistics
may suffer the problem to keep the time reversal
symmetry. To keep it, one may have to use a $ Z_2 $ CS term on a lattice as did in \cite{SF}.
For a review on this connection, see Ref.\cite{wenrev,SLrev3} and references therein.


\bibitem{SF} T. Senthil and Matthew P. A. Fisher, Z2
 gauge theory of electron fractionalization in strongly correlated systems,
Phys. Rev. B 62, 7850 每 Published 15 September 2000

\bibitem{Kit2} A. Kitaev, Anyons in an exactly solved model and beyond.
Ann. Phys. 321, 2每111 (2006). doi: 10.1016/j.aop.2005.10.005

\bibitem{HKmodel} G. Jackeli, G. Khaliullin, Mott insulators in the strong spinorbit
coupling limit: From Heisenberg to a quantum
compass and Kitaev models. Phys. Rev. Lett. 102, 017205 (2009).

\bibitem{SLrev1} L. Savary and L. Balents, \textit{Quantum Spin liquids},
arXiv:1601.03742 (2016).


\bibitem{SLrev3} C. Broholm1, R. J. Cava, S. A. Kivelson, D. G. Nocera, M. R. Norman, T. Senthil,
Quantum spin liquids, Science  17 Jan 2020: Vol. 367, Issue 6475, eaay0668, DOI: 10.1126/science.aay0668



\bibitem{CSL} In this work, we only focused on gapped $ Z_2 $ QSL which breaks no symmetries at all.
There could also be a chiral QSL (CSL) which breaks
the Time reversal and parity, but nothing else, first proposed by Kalmeyer and Laughlin \cite{KL}.
The original Abelian CSL is essentially the same as the $ \nu=1/2 $ bosonic fractional quantum Hall states.
The CSL was shown to be an exact ground state in the extended Kitaev model in the honeycomb-triangular lattice \cite{dimer3}.
It also hosts both Abelian and non-Abelian excitations separated by a topological phase transition.
DMRG \cite{SLrev1} also suggests the Abelian CSL may be the ground state of a Kagome lattice with $ J_1-J_2-J_3 $ spin exchange interactions.

\bibitem{KL} V. Kalmeyer, R. B. Laughlin, Equivalence of the resonatingvalence-
bond and fractional quantum Hall states. Phys. Rev. Lett. 59, 2095每2098 (1987).


\bibitem{rh} Fadi Sun, Jinwu Ye, Wu-Ming Liu, Rotated Heisenberg Model, Phys. Rev. A 92, 043609 (2015).

\bibitem{rhht} Fadi Sun,  Jinwu Ye and Wu-Ming Liu, Classification of magnons in Rotated Ferromagnetic Heisenberg model and their competing responses in transverse fields, Phys. Rev. B 94, 024409 ( 2016 ).

\bibitem{rafhm}  Fa-Di Sun and Jinwu Ye and Wu-Ming Liu, Fermionic Hubbard model with Rashba or Dresselhaus spin每orbit coupling,  New J. Phys. 19, 063025 (2017).

\bibitem{rhh}  Fadi Sun, Jinwu Ye, Wu-Ming Liu, Quantum incommensurate skyrmion crystals and commensurate to in-commensurate transitions in cold atoms and materials with spin每orbit couplings in a Zeeman field, New J. Phys. 19, 083015 (2017).

\bibitem{devil} Fadi Sun and Jinwu Ye,
 In-complete, complete devil staircases and Luttinger liquids Cantor set of
 strongly interacting spin-orbit coupled bosons in a square lattice,
 arXiv:1603.00451, substantially revised version.




\bibitem{gold} Fadi Sun and Jinwu Ye,
Goldstone modes generated by order from quantum disorder and its experimental observation in cold atom or photonic systems, Preprint.
See also therein for the list of recent cold atom and photonic experiments generating an Abelian flux.



\bibitem{kane} M. Z. Hasan and C. L. Kane, Colloquium: Topological insulators,  Rev. Mod. Phys. {\bf 82}, 3045 (2010).

\bibitem{zhang} X. L. Qi and S. C. Zhang, Topological insulators and superconductors, Rev. Mod. Phys. {\bf 83}, 1057 (2011).

\bibitem{tenfold} Ching-Kai Chiu, Jeffrey C.?Y. Teo, Andreas P. Schnyder, and Shinsei Ryu, Classification of topological quantum matter with symmetries, Rev. Mod. Phys. 88, 035005 (2016) - Published 31 August 2016.

\bibitem{wenrev} 
Xiao-Gang Wen, Colloquium: Zoo of quantum-topological phases of matter,
		Rev. Mod. Phys. {\bf 89}, 041004 (2017).

\bibitem{etafluc}    To be complete, one need also consider the quantum fluctuations in the two sub-lattice eigen-vectors $ \eta_{1,2} $,
    but these are high energy modes which can can be dropped in constructing  low energy effective action.




\bibitem{SY} S. Sachdev and J. Ye,  Gapless spin-fluid ground state in a random quantum Heisenberg magnet,
   Phys. Rev. Lett. 70, 3339 (1993).









\bibitem{pq1}
L. Balents, L. Bartosch, A. Burkov, S. Sachdev, and K. Sengupta,
\emph{Putting competing orders in their place near the Mott transition},
Phy. Rev. B {\bf 71}, 144508 (2005).


\bibitem{yan}
Yan Chen and Jinwu Ye,
``Characterizing boson orders in lattices by vortex degree of freedoms'',
Philos. Mag. 92, 4484 (2012);
``Quantum phases, Supersolids and quantum phase transitions of interacting bosons in frustrated lattices'',
Nucl. Phys. B 869, 242 (2013).




























\bibitem{alsonoflux}
In fact, in the two component bosons Eq.\ref{piflux}
in the absence of any flux, there is only one minimum in the kinetic energy,  one can get an effective action
identical to Eq.\ref{singlecomponent}. So it also describes effectively the FM-SF to SF-Mott transition in this case.
The case of one component in the presence of $ \pi $ flux will be discussed in the Appendix C.




\bibitem{tunneling}  Based on the similarity between the $ \epsilon $ particle in the QSL and that in the type-II superconductors \cite{myown},
    possible sensitive tunneling experimental detections of these fractionalized particles $ e,m, \epsilon $ between
    the two systems  have been proposed in \cite{stevensci}, but not implemented yet.


\bibitem{softQMC}
 F. Hebert {\sl  et.al},  Phys. Rev. B {\bf 65}, 014513 (2001);
 P. Sengupta, {\sl et.al}, Phys. Rev. Lett. {\bf 94}, 207202 (2005).
 Jing Yu Gan, Yu Chuan Wen, Jinwu Ye, Tao Li, Shi-Jie Yang, Yue Yu, Phys. Rev. B 75, 214509 (2007);  S. Wessel, Phys. Rev. B 75, 174301 (2007)

\bibitem{JQ}
Our J-Q  model is dramatically different than the one proposed by Sandvick  which shows a deconfined
QCP from a Neel state to a VBS. See,  Anders W. Sandvik, Phys. Rev. Lett. 98, 227202 (2007).
Note that Sandvick's J-Q model can not model the strong coupling expansion of the fermionic Hubbard model.
For example,  to the order of $t^4/U $,  it dropped many terms in the strong coupling expansion, keep only the two terms
where $ ij $ and $ k l $ form two parallel adjacent horizontal or vertical links. Most importantly, it
also change the sign of the two terms to make it QMC sign free. So it is a specially
designed sign-free version of J-Q model with $ J >0, Q < 0 $.



\bibitem{braggbog} M. Kozuma, {\sl et.al}, Phys. Rev. Lett. 82, 871 (1999); J. Stenger, {\sl et al}, Phys. Rev. Lett. 82, 4569 (1999);
D. M. Stamper-Kurn {\sl et al},
   Phys. Rev. Lett. 83, 2876 - 2879 (1999); J. Steinhauer, {\sl et.al}, Phys. Rev. Lett. 88,
   120407, (2002); S. B. Papp, {\sl et.al}, Phys. Rev. Lett. 101, 135301 (2008)

\bibitem{braggangle}   P. T. Ernst, {\sl et al}, Nature Physics 6, 56 (2010 ).

\bibitem{braggeng}  T. Stoferle {\sl et al}, Phys. Rev. Lett. 92, 130403 (2004).



\bibitem{braggsingle}  G. Birkl, {\sl et al},  Phys. Rev. Lett. 75, 2823
(1995); M. Weidem邦ller, {\sl et al},  Phys. Rev. Lett. 75, 4583
(1995), Phys. Rev. A 58, 4647 (1998).
 J. Ruostekoski, C. J. Foot, and A. B. Deb, Phys. Rev. Lett. 103,
170404 (2009).

\bibitem{becbragg}  Si-Cong Ji, Long Zhang, Xiao-Tian Xu, Zhan Wu, Youjin Deng, Shuai Chen, and Jian-Wei Pan, Phys. Rev. Lett. 114, 105301 每 Published 9 March 2015,

\bibitem{bragg12}  Jinwu Ye, J.M. Zhang, W.M. Liu, K.Y. Zhang, Yan Li, W.P. Zhang, Phys. Rev. A 83, 051604 (R) (2011);
Jinwu Ye, K.Y. Zhang, Yan Li, Yan Chen and W.P. Zhang, Ann. Phys. 328 (2013), 103-138.


\bibitem{stevekondo}
V. J. Emery and S. Kivelson,
\emph{Mapping of the two-channel Kondo problem to a resonant-level model},
Phys. Rev. B 46, 10812 (1992).

\bibitem{Malkondo}
Juan M. Maldacena, Andreas W. W. Ludwig, Majorana Fermions,
\emph{Exact Mapping between Quantum Impurity Fixed Points with four bulk Fermion species, and Solution of the ``Unitarity Puzzle''},
Nucl.Phys. B506 (1997) 565-588.

\bibitem{kondoye12345}
Jinwu Ye,
\emph{On Emery-Kivelson line and universality of Wilson ratio of spin anisotropic Kondo model},
Phys. Rev. Lett. 77, 3224 (1996).

\bibitem{kondoye123452}
Jinwu Ye,
\emph{Abelian Bosonization approach to quantum impurity problems},
Phys. Rev. Lett. 79, 1385 (1997).

\bibitem{QSG1} J. Ye, S. Sachdev and N. Read, A solvable spin glass of quantum rotors,
    Phys. Rev. Lett. 70, 4011 (1993).

\bibitem{QSG2}  N. Read, S. Sachdev and J. Ye, Landau theory of quantum spin glasses of rotors and Ising spins,
 Phys.Rev.B, 52,  384 (1995).

\bibitem{stevensci}
Maissam Barkeshli, Erez Berg, Steven Kivelson, Coherent transmutation of electrons into fractionalized anyons,
Science  07 Nov 2014, Vol. 346, Issue 6210, pp. 722-725,
DOI: 10.1126/science.1253251


\bibitem{senthil} T. Senthil and Michael Levin, Integer Quantum Hall Effect for Bosons, Phys. Rev. Lett. 110, 046801 每 Published 22 January 2013;
Ashvin Vishwanath and T. Senthil, Physics of Three-Dimensional Bosonic Topological Insulators: Surface-Deconfined Criticality and Quantized Magnetoelectric Effect, Phys. Rev. X 3, 011016 每 Published 28 February 2013;
T. Senthil and Matthew P. A. Fisher, Competing orders, nonlinear sigma models, and topological terms in quantum magnets, Phys. Rev. B 74, 064405 每 Published 8 August 2006.



\bibitem{xu} Cenke Xu and Andreas W. W. Ludwig, Nonperturbative Effects of a Topological Theta Term on Principal Chiral Nonlinear Sigma Models in
$ 2+1 $ Dimensions, Phys. Rev. Lett. 110, 200405 每 Published 17 May 2013

\bibitem{max} Max A. Metlitski, C. L. Kane, and Matthew P. A. Fisher,
Bosonic topological insulator in three dimensions and the statistical Witten effect,
Phys. Rev. B 88, 035131 (2013) - Published 25 July 2013

\bibitem{decon}  T. Senthil, Leon Balents, Subir Sachdev, Ashvin Vishwanath, and Matthew P. A. Fisher,
Quantum criticality beyond the Landau-Ginzburg-Wilson paradigm,
Phys. Rev. B 70, 144407 (2004) - Published 15 October 2004

\bibitem{frustri} There are also two possible ways to frustrate the 3 sublattice $ 120^{\circ} $ coplanar
spin state in a triangular lattice \cite{SLrev1}: (1) the $ J_1-J_2 $ model, the NNN $ J_2 $ term may melt it into a $ Z_2 $  QSL.
(2) The Hubbard model in a triangular lattice. The ring exchange term close to the metallic state may melt it into
a $ U(1) $ gapless QSL with a spinon FS coupled to a $ U(1)  $ gauge field.
However, due to the absence of exact solutions for this kind of strongly coupled gapless $ U(1) $ QSL, its nature, even its stability
at 2d is not known yet.

\bibitem{renyi}  Rajibul Islam, Ruichao Ma, Philipp M. Preiss, M. Eric Tai, Alexander Lukin, Matthew Rispoli, Markus Greiner,
Measuring entanglement entropy through the interference of quantum many-body twins, arXiv:1509.01160 (cond-mat).


\bibitem{bloch1} Jayadev Vijayan, Pimonpan Sompet, Guillaume Salomon, Joannis Koepsell, Sarah Hirthe, Annabelle Bohrdt, Fabian Grusdt, Immanuel Bloch, Christian Gross, Time-Resolved Observation of Spin-Charge Deconfinement in Fermionic Hubbard Chains, Science 367, 186 (2020).

\bibitem{bloch2} Christian Schweizer, Fabian Grusdt, Moritz Berngruber, Luca Barbiero, Eugene Demler, Nathan Goldman, Immanuel Bloch, Monika Aidelsburger,  Floquet approach to Z2 lattice gauge theories with ultracold atoms in optical lattices, Nature Physics 15, 1168-1173 (2019).

\bibitem{bloch3}
 Guillaume Salomon, Joannis Koepsell, Jayadev Vijayan, Timon A. Hilker, Jacopo Nespolo, Lode Pollet, Immanuel Bloch, Christian Gross,
 Direct observation of incommensurate magnetism in Hubbard chains, Nature (2018).

\bibitem{blqhye}  Jinwu Ye, Mutual Composite Fermion and Composite Boson approaches to balanced and im-balanced bilayer quantum Hall
       systems: an electronic analogy of Helium 4 system,  Annals of Physics, 323 (2008), 580-630.

\bibitem{fermion} The gap closing of the $\epsilon $ particle could lead to a gapless $ U(1) $ QSL \cite{frustri}.
   This scenario is unlikely in the present case.

\end{thebibliography}
\end{document}